\documentclass[preprint,12pt]{elsarticle}

\usepackage{amssymb}
\usepackage{amsmath}
\usepackage{booktabs}
\usepackage{hyperref} 
\usepackage{tabularx}
\usepackage{array}
\usepackage{graphicx}
 \usepackage{longtable} 
\usepackage{subcaption}
\usepackage{pdfpages} 
\usepackage[gen]{eurosym}
\usepackage{xcolor}

\usepackage{url}
\usepackage{lscape}
\usepackage[superscript,biblabel]{cite}
\makeatletter
\def\@biblabel#1{\hbox{#1.}}
\makeatother

\begin{document}
\begin{frontmatter}


\title{Association of Timing and Duration of Moderate-to-Vigorous Physical Activity with Cognitive Function and Brain Aging: A Population-Based Study Using the UK Biobank}

\author[1]{Wasif Khan}
\author[2]{Lin Gu}
\author[3]{Noah Hammarlund}
\author[4]{Lei Xing}
\author[5]{Joshua K. Wong}

\author[1,7,9]{Ruogu Fang\corref{Ruogu Fang}}


\affiliation[1]{
    organization={J. Crayton Pruitt Family Department of Biomedical Engineering, University of Florida},
    city={Gainesville},
    state={FL},
    country={USA}
}

\affiliation[2]{
    organization={RIKEN},
    country={Japan}
}

\affiliation[3]{
    organization={Department of Health Services Research, Management and Policy, University of Florida},
    city={Gainesville},
    state={FL},
    country={USA}
}

\affiliation[4]{
    organization={Department of Radiation Oncology, Stanford University},
    state={CA},
    country={USA}
}

\affiliation[5]{
    organization={Norman Fixel Institute for Neurological Diseases, University of Florida},
    city={Gainesville},
    state={FL},
    country={USA}
}

\affiliation[7]{
    organization={Center for Cognitive Aging and Memory, McKnight Brain Institute, University of Florida},
    city={Gainesville},
    state={FL},
    country={USA}
}
\affiliation[9]{
    organization={Department of Electrical and Computer Engineering, University of Florida},
    city={Gainesville},
    state={FL},
    country={USA}
}

\cortext[Ruogu Fang]{Corresponding author}
\ead{ruogu.fang@bme.ufl.edu}

\begin{abstract}

\noindent\textbf{Background:}
Physical activity is a modifiable lifestyle factor with potential to support cognitive resilience. However, the association of moderate-to-vigorous physical activity (MVPA) intensity, and timing, with cognitive function and region-specific brain structure remain poorly understood. \\
\textbf{Methods:}
We analyzed data from 45,892 UK Biobank participants aged 60 years and older with valid wrist-worn accelerometer data, cognitive testing, and structural brain MRI. MVPA was measured both continuously (mins/week) and categorically (thresholded using $\geq$150 min/week based on WHO guidelines). Associations with cognitive performance and regional brain volumes were evaluated using multivariable linear models adjusted for demographic, socioeconomic, and health-related covariates. We conducted secondary analyses on MVPA timing and subgroup effects. \\
\textbf{Findings:} 
Higher MVPA was associated with better performance across cognitive domains, including reasoning, memory, executive function, and processing speed. These associations persisted in fully adjusted models and were higher among participants meeting WHO guidelines. Greater MVPA was also associated with subcortical brain regions (caudate, putamen, pallidum, thalamus), as well as regional gray matter volumes involved in emotion (insula), working memory (cerebellar lobules VI, Crus I, VIIIa), and perceptual processing (fusiform gyrus). Secondary analyses showed that MVPA at any time of day was associated with cognitive functions and brain volume particularly in the midday-afternoon and evening. Sensitivity analysis shows consistent findings across subgroups, with evidence of dose–response relationships.\\
\textbf{Interpretation:}
Higher MVPA is associated with preserved brain structure and enhanced cognitive function in later life. Public health strategies to increase MVPA may support healthy cognitive aging and generate substantial economic benefits, with global gains projected to reach USD 760 billion annually by 2050.

\end{abstract}


\end{frontmatter}


\section{Introduction}

As populations age worldwide, the preservation of cognitive health has become a pressing public health priority with far-reaching social and economic consequences~\cite{iso2024physical}. Cognitive decline from mild cognitive impairment to dementia can significantly affect quality of life, independence, healthcare costs, and overall health in aging populations \cite{livingston2020dementia}. In the absence of curative measures, promoting lifestyle factors can improve cognitive resilience \cite{song2022modifiable,brodaty2025online} and support healthy aging \cite{ fox2022association, hofman2023physical, melo2023objective},
with higher physical activity (PA) increases the odds of healthy ageing by 39\%~\cite{daskalopoulou2017physical}.

The Lancet commission~\cite{livingston2020dementia} identifies physical inactivity as a potential modifiable risk factor for cognitive decline, and dementia. Furthermore, it estimates that up to 45\% of dementia cases could be prevented by addressing 14 modifiable risk factors~\cite{livingston2024dementia}. Global health guidelines such as the World Health Organization (WHO) recommend for all adults engaging in at least 150–300 minutes of moderate-intensity, or 75–150 minutes of vigorous-intensity aerobic activity per week~\cite{livingston2020dementia,world2024nearly}. Regular PA supports cardiovascular, metabolic health, improves mood, reduces anxiety, maintain cognitive performance in later life, and is increasingly recognized as beneficial for brain health~\cite{voss2013neurobiological,hamer2018association}. Moreover, low-to-moderate and higher level of PA reduce the risk of cognitive decline by 35\% and 38\%, respectively~\cite{sofi2011physical}. 

Recent studies show the association of PA, cognitive function, and brain volume, however, findings remain mixed \cite{hofman2023physical,melo2023objective,hamer2018association,sabia2017physical,campbell2023estimating, zhu2017objectively,feter2021physical,groot2016effect,erickson2011exercise}. For instance, Zhu et. al show that higher MVPA quartiles are associated with 36\% lower odds of cognitive impairment in older adults\cite{zhu2017objectively}. Longitudinal cohort data, such as from the U.S. Health and Retirement Study, show that regular PA is associated with 
with a 30–49\% reduced risk of dementia and up to 23\% reduced risk of cognitive impairment over 12 years of follow-up~\cite{wei2024physical}. Furthermore, while some studies show the association of light-intensity physical activity (LPA) and cognition~\cite{melo2023objective,spartano2019association}; for instance, the Framingham Heart Study found associations between LPA and greater total brain volume, equivalent to approximately 1.1 years of delayed brain aging~\cite{spartano2019association}. However, other studies found no significance association of LPA with cognition\cite{zhu2017objectively,erlenbach2021association}. Moreover, studies utilizing large-scale data such as Whitehall II and UK Biobank (UKB), including controlled trials show limited or null effects on cognitive decline or mortality~\cite{sabia2017physical, campbell2023estimating, brown2021high}.

Despite valuable insights, prior studies have notable limitations such as relied on self-reported PA~\cite{hofman2023physical,sabia2017physical,feter2021physical,wei2024physical} which is prone to recall bias~\cite{min2024accelerometer}, daily acceleration which is difficult to contextualize for PA recommendation \cite{campbell2023estimating},  small sample size~\cite{melo2023objective,erickson2011exercise}, limited focus on older adults~\cite{hamer2018association,sabia2017physical}, and overlooked to examine MVPA intensity and timing patterns with both cognitive function and brain regions. Furthermore, recent studies also focus on dementia incidence, restricting their analyses to individuals who later develop dementia.
To address these limitations, we utilized accelerometer-derived activity data, detailed cognitive assessments, and multimodal brain MRI from the UKB to examine the association of MVPA with primary outcomes: cognitive function and brain volumes in older adults. Our approach captures variation in cognitive and brain health across the general older population, without restricting to those who later develop dementia. Specifically, we examined the association of MVPA measured both continuously (mins/week) and categorically (thresholded using $\geq$150 min/week based on WHO guidelines) with cognitive performance, subcortical regions and regional grey matter brain volumes. Furthermore, we investigated how MVPA intensity and timing patterns related to cognitive and neuroimaging outcomes. We conducted sensitivity analyses to explore potential dose–response relationships, and performed subgroup analyses stratified by age, sex, and obesity. In secondary analyses, we investigated associations between MVPA timing with cognitive performance and brain volumes. 


\begin{figure}
    \centering
    \includegraphics[width=1\linewidth, trim=0cm 1cm 0cm 1cm, clip]{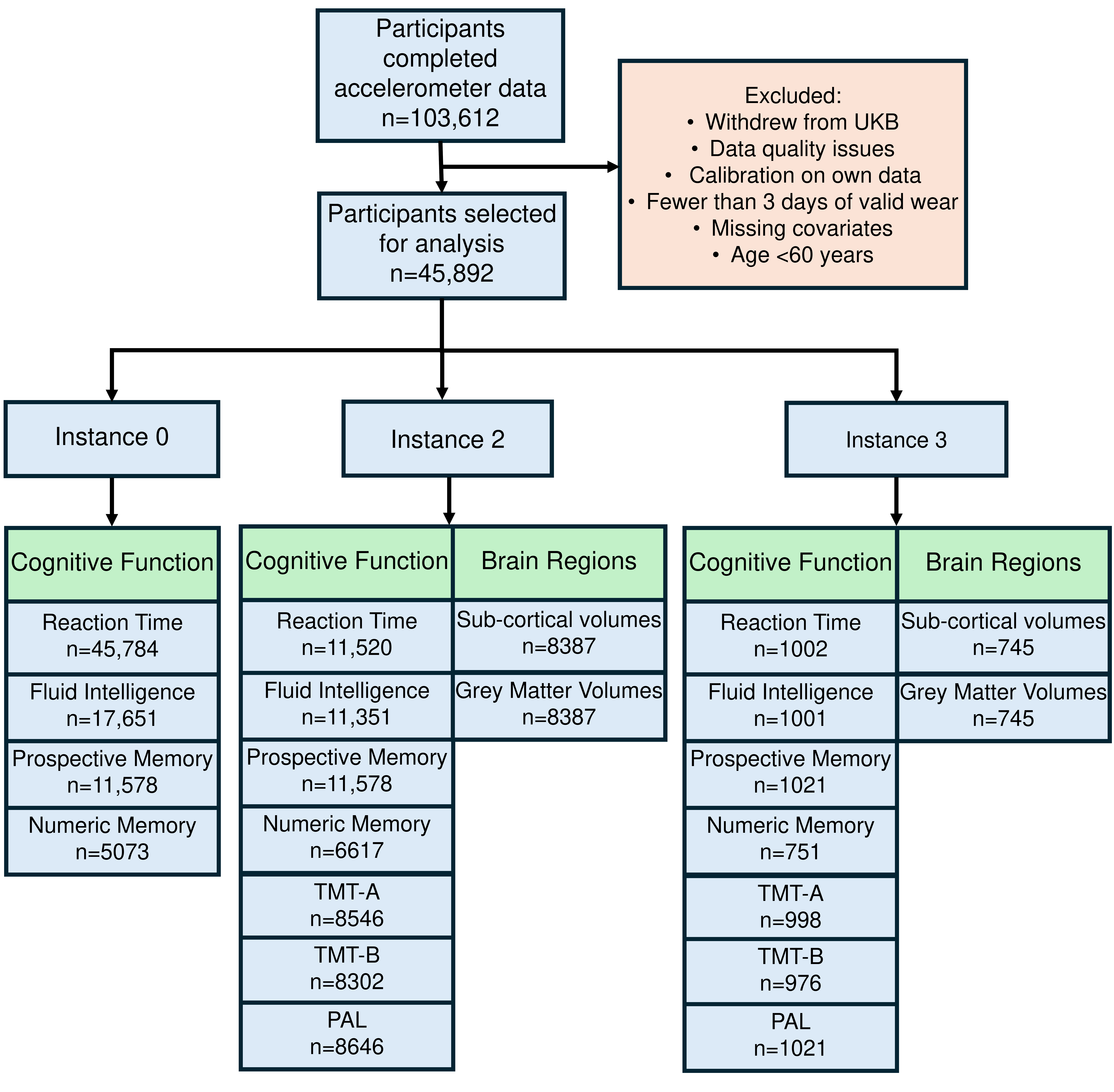}
    \caption{ Flowchart of participants selection and outcome availability across UK Biobank assessment instances. From 103,612 participants with valid wrist-worn accelerometer data, 45,892 individuals aged $\geq$60 years were retained for analysis after exclusion criteria. Cognitive function and brain MRI data were obtained across three UK Biobank imaging and cognitive assessment instances (Instance 0, 2, and 3), with sample sizes varying depending on availability and completion of specific outcomes. TMT-A: trail making test A (numeric), TMTB: Trail making test B (alphanumeric), PAL: Paired associate learning.}
    \label{fig:enter-label}
\end{figure}

{\small
\begin{longtable}{p{3.8cm}p{2.5cm}p{2.8cm}p{2.8cm}}
\caption{Descriptive characteristics of the study population stratified by weekly moderate-to-vigorous physical activity (MVPA) levels. Participants are grouped into those meeting the World Health Organization (WHO) guidelines (\(\geq150\) minutes/week) and those engaging in less than 150 minutes/week of MVPA. Cognitive function was assessed at three timepoints in the UK Biobank: baseline assessment (instance 0), imaging visit (instance 2), and imaging follow-up (instance 3). Arrows indicate direction of favorable outcomes: $\uparrow$: higher values indicate superior cognitive performance; $\downarrow$: lower values indicate faster or more efficient performance.} \\
\toprule
\textbf{Characteristic} & \textbf{Total} & \multicolumn{2}{c}{\textbf{WHO Guidelines}} \\
& (n=45,892) & \text{$\geq$150}min/week (n=29,755) & \textless150 min/week (n=16,137) \\
\midrule
\endfirsthead
\toprule
\textbf{Characteristic} & \textbf{Total} & \multicolumn{2}{c}{\textbf{WHO Guidelines}} \\
& (n=45,892) & \text{$\geq$150}min/week  (n=29,755) & \textless150 min/week (n=16,137) \\
\midrule
\endhead
Age & 66.97 ± 4.16 & 67 ± 4 & 67 ± 4 \\
\midrule
Sex & & & \\
Female & 24,380 (53\%) & 14,114 (47\%) & 10,266 (64\%) \\
Male & 21,512 (47\%) & 15,641 (53\%) & 5,871 (36\%) \\
\midrule
Ethnicity & & & \\
Others & 990 (2.2\%) & 581 (2.0\%) & 409 (2.5\%) \\
White & 44,902 (98\%) & 29,174 (98\%) & 15,728 (97\%) \\
\midrule
BMI & 26.8 ± 4.4 & 26.1 ± 3.8 & 28.1 ± 5.0 \\
\midrule
Systolic blood pressure (mmHg) & 143 ± 19 & 142 ± 19 & 143 ± 19 \\
\midrule
Diastolic blood pressure (mmHg) & 82 ± 10 & 82 ± 10 & 82 ± 10 \\
\midrule
HDL cholesterol (mmol/l) & 1.50 ± 0.39 & 1.52 ± 0.40 & 1.47 ± 0.39 \\
\midrule
LDL direct (mmol/l) & 3.63 ± 0.87 & 3.63 ± 0.85 & 3.62 ± 0.91 \\
\midrule
Triglycerides (mmol/l) & 1.72 ± 0.94 & 1.67 ± 0.91 & 1.83 ± 0.98 \\
\midrule
Smoking & & & \\
Never* & 24,639 (54\%) & 16,397 (55\%) & 8,242 (51\%) \\
Previous & 18,646 (41\%) & 11,938 (40\%) & 6,708 (42\%) \\
Current & 2,607 (5.7\%) & 1,420 (4.8\%) & 1,187 (7.4\%) \\
\midrule
Alcohol use & & & \\
Never* & 2,656 (5.8\%) & 1,444 (4.9\%) & 1,212 (7.5\%) \\
Daily & 12,126 (26\%) & 8,462 (28\%) & 3,664 (23\%) \\
1–4 times/week & 22,421 (49\%) & 15,029 (51\%) & 7,392 (46\%) \\
Occasionally & 8,689 (19\%) & 4,820 (16\%) & 3,869 (24\%) \\
\midrule
Townsend D Index & -1.93 ± 2.68 & -1.93 ± 2.68 & -1.94 ± 2.68 \\
\midrule
Diabetes History & 1,977 (4.3\%) & 899 (3.0\%) & 1,078 (6.7\%) \\
\midrule
Level of Education & & & \\
Never* & 5,464 (12\%) & 2,967 (10\%) & 2,497 (15\%) \\
A level, O level, or equivalent & 22,022 (48\%) & 13,523 (45\%) & 8,499 (53\%) \\
College/University & 18,406 (40\%) & 13,265 (45\%) & 5,141 (32\%) \\
\midrule
Longstanding Illness & 14,561 (32\%) & 8,099 (27\%) & 6,462 (40\%) \\
\midrule
Reaction Time $\downarrow$ &  &  &  \\
Instance 0 & 564 ± 109 & 561 ± 108 & 570 ± 110 \\
Instance 2 & 621 ± 113 & 618 ± 112 & 629 ± 115 \\
Instance 3 & 624 ± 112 & 621 ± 106 & 632 ± 127 \\
\midrule
Fluid Intelligence  $\uparrow$ &   &  &  \\
Instance 0 & 6.54 ± 2.03 & 6.65 ± 2.03 & 6.34 ± 1.99 \\
Instance 2 & 6.60 ± 2.02 & 6.68 ± 2.03 & 6.42 ± 1.97 \\
Instance  3 & 6.69 ± 2.00 & 6.81 ± 1.94 & 6.38 ± 2.11 \\
\midrule
\multicolumn{2}{l}{Duration Numeric Path trail $\downarrow$}  &  &  \\
Instance 2 & 248 ± 93 & 247 ± 95 & 248 ± 89 \\
Instance 3 & 243 ± 108 & 242 ± 118 & 246 ± 80 \\
\midrule
\multicolumn{2}{l}{Duration Alpha-Numeric Path Trail $\downarrow$}&  & \\
Instance 2 & 639 ± 283 & 632 ± 282 & 657 ± 284 \\
Instance 3 & 614 ± 291 & 608 ± 294 & 630 ± 283 \\
\midrule
\multicolumn{2}{l}{Number Word Pairs Correct $\uparrow$}&  & \\
Instance 2 & 6.49 ± 2.64 & 6.56 ± 2.61 & 6.33 ± 2.72 \\
Instance 3 & 6.76 ± 2.65 & 6.89 ± 2.65 & 6.45 ± 2.62 \\
\midrule
\multicolumn{2}{l}{Max Digit Remembered Correctly $\uparrow$}&  & \\
Instance 0 & 6.86 ± 1.23 & 6.92 ± 1.22 & 6.73 ± 1.25 \\
Instance 2 & 6.66 ± 1.27 & 6.69 ± 1.27 & 6.58 ± 1.28 \\
Instance 3 & 6.75 ± 1.23 & 6.79 ± 1.20 & \\
\midrule
Prospective Memory $\uparrow$  First-attempt recall & & & \\
Instance 0& 14,734 (83\%) & 9,640 (83\%) & 5,094 (82\%) \\
Instance 2 & 9,369 (81\%) & 6,620 (81\%) & 2,749 (80\%) \\
Instance 3 & 863 (85\%) & 624 (86\%) & 239 (80\%) \\
\bottomrule
\end{longtable}}
\vspace{-0.5em}
*Responses of “Prefer not to answer,” “Never,” and “None of the above” were grouped into a single category.


\section{Methods}

\subsection*{Study design and participants}
We used UKB data from 103,612 participants who completed 7-day wrist-worn accelerometer between 2013 and 2015~\cite{doherty2017large}. MVPA was estimated using Random Forest~\cite{walmsley2022reallocation}, and analyzed both as a continuous variable (minutes/week) and categorical based on WHO guidelines ($\geq$150 min/week). Participants aged $\geq$60 years with valid accelerometer and complete covariates were included (N = 45,892; appendix pp 3).

\subsection*{Cognitive Function and Brain Imaging}
Cognitive outcomes were derived from UKB touchscreen assessments conducted at baseline (Instance 0), first imaging visit (Instance 2), and first repeat imaging visit (Instance 3). Primary measures included reaction time, fluid intelligence, prospective and numeric memory, trail making tests A and B (TMT-A and TMT-B), and paired associate learning (appendix Table S.1). We also analyzed two primary outcomes T1-weighted MRI processed by UKB. (i) Subcortical volumes of 14 bilateral structures (hippocampus, amygdala, thalamus, caudate, putamen, pallidum, and nucleus accumben) were estimated using FMRIB’s Integrated Registration and Segmentation Tool (FIRST)~~\cite{alfaro2018image}. Regional grey matter volumes across 139 cortical and subcortical regions were derived using FMRIB’s Automated Segmentation Tool (FAST) to assess localized structural variation~\cite{alfaro2018image}.


\subsection*{Covariates}
Covariates were selected based on their established associations with both PA and cognitive function in older adults~\cite{taylor2024association, feng2023associations}. Age was calculated as participant’s date of birth and the date of wearing accelerometer, quadratic age ($\text{age}^2/1000$) to account for potential non-linear age-related effects~\cite{taylor2024association}. Demographic factors included sex (acquired from central registry at recruitment), ethnicity, smoking status, alcohol consumption frequency, educational attainment, and long-term illness were all assessed via self-report at the assessment center. Townsend Deprivation Index (TDI), a continuous score derived from residential postcode data reflecting area-level deprivation. Health-related and biomedical covariates included body mass index (BMI), systolic and diastolic blood pressure (SBP and DBP), triglyceride levels, high-density (HDL) and low-density (LDL) lipoprotein cholesterol, and history of diabetes, all of which were measured during clinical assessments or retrieved from linked health records. PA exposures were derived from accelerometer data. History of cancer was recorded via self-report and confirmed with registry data. Detailed information on the covariates can be found in Supplementary Table S.2.

\subsection*{Statistical Analysis}
All statistical analyses were performed using R using the UKB-RAP platform. Descriptive statistics were used to summarize the study sample. Continuous variables were summarized using means and standard deviations, while categorical variables were summarized as frequencies and percentages. Multivariable linear regression models were used for association between MVPA and cognitive function and brain volumes. We used two models based on covariates adjustments (1) based model adjusted age and sex, (2) fully adjusted model adjusted for age, age with quadratic term ($\text{age}^2/1000$), sex, ethnicity, smoking status, alcohol consumption frequency, educational attainment, long-term illness, TDI, HDL cholestrol, LDL direct, DBP, SBP, Triglycerides, longstanding illnes,BMI, and  diabetes history. MVPA was evaluated both as a continuous variable (minutes per week) and categorized based on WHO recommendations ($\geq$150 min/week). For neuroimaging analyses, linear models were similarly applied to assess associations between MVPA and subcortical structures (FIRST) and regional grey matter volumes (FAST). Participants with missing data on any covariate or outcome were excluded and all models included complete case analysis. The analysis were performed for each visit (Instance 0, 2, 3) separately. Sensitivity analyses tested MVPA dose–response effects and stratified models by age, sex, and obesity (excluding the stratifying variable from covariates). In secondary analyses, we categorized MVPA timing into Morning, Afternoon, Evening, and Mixed profiles based on diurnal patterns. Fully adjusted linear models examined associations between MVPA timing and cognitive outcomes, subcortical volumes, and regional grey matter volumes to assess potential timing-specific effects on brain health.


\section{Results}
Our experiments includes 45,892 participants(Figure 1). Descriptive statistics, stratified by weekly MVPA levels based on the WHO guidelines, are presented in Table 1. Participants meeting the guideline (N = 29,755) had lower BMI (26.1 compared to 28.1 for individual with MVPA less than 150 mins/week), higher education levels, and lower history of diabetes prevalence. They also showed slightly better cognitive performance: faster reaction time, memory, and reasoning—compared to those below the threshold (N = 16,137). The distribution of the cognitive variables is shown in Supplementary Figure S.1. The results in below sections reports p-values when the associations are statistically significant (\( p<0.05 \)). 

\subsection{Cognitive Function}
\begin{figure}[htbp]
    \centering
    \begin{subfigure}[t]{\linewidth}
        \centering
        \includegraphics[width=1\linewidth]{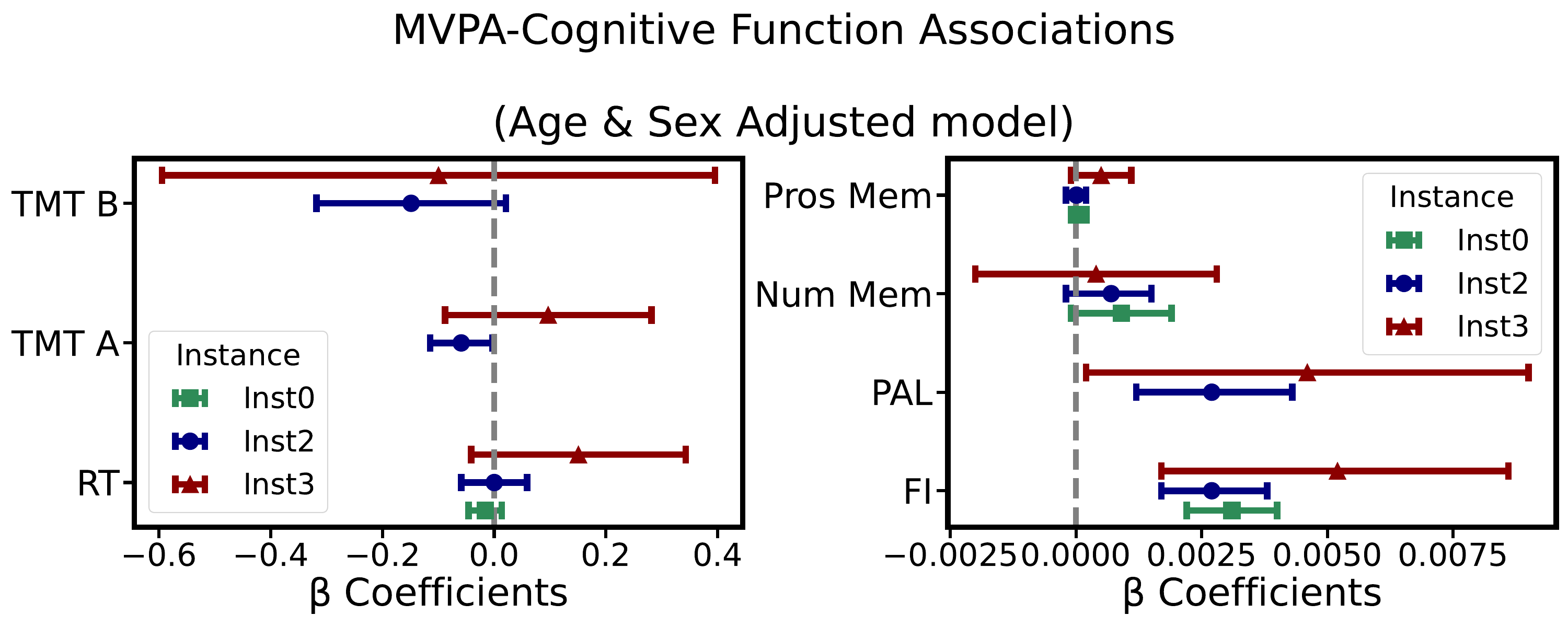}
    \end{subfigure}


    \begin{subfigure}[t]{\linewidth}
        \centering
        \includegraphics[width=1\linewidth]{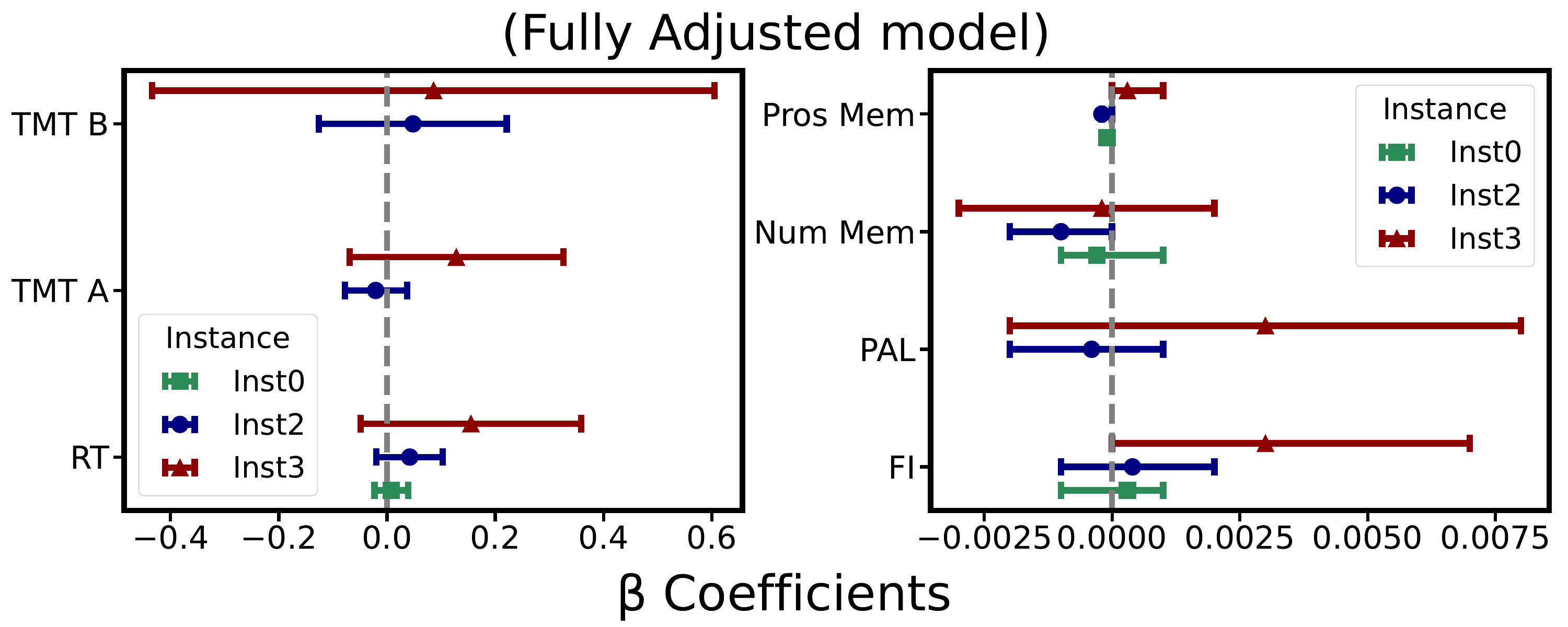}
        \label{fig:mvpa_full_adj}
    \end{subfigure}

    \caption{Associations between MVPA and cognitive function across multiple domains. Results are shown for both based ( age-sex adjusted) and fully adjusted models. Abbreviations: RT = Reaction Time, Inst = Instance, Fluid = Fluid Intelligence, TMT-A = Trail Making Test A, TMT-B = Trail Making Test B, PAL = Paired Associate Learning, NumMem = Numeric Memory, ProsMem = Prospective Memory, Inst0= Instance 0, Inst2=Instance 2, Inst3=Instance 3.}
    \label{fig:mvpa_cognition_combined}
\end{figure}

In the base model adjusted for age and sex, MVPA was associated with better cognitive function across several domains for both continuous (minutes/week) and categorical (WHO guideline-based) MVPA metrics (Table~S3). Continuous MVPA (Figure \ref{fig:mvpa_cognition_combined}) was associated with higher fluid intelligence  at baseline ($\beta=0.0031$, $p<0.0001$); Instance~2 ($\beta=0.0027$, $p<0.0001$) ; and Instance~3 ($\beta=0.0052$, $p=0.0031$). Similarly, at instance 2, MVPA was associated with better paired associate learning ($\beta=0.0027$, $p=0.0006$), and  Trail Making Test (TMT)-A ($\beta=-0.0589$, $p=0.038$). These findings suggest that individuals with higher MVPA perform better on tasks involving reasoning, memory, and attention-switching. Stronger associations where shows for categorical MVPA (WHO guidelines-based) (Table~S3). Participants meeting the guidelines showed higher fluid intelligence scores at Instance 0 ($\beta=0.2454$, $p<0.0001$); Instance 2 ($\beta=0.191$, $p<0.0001$) and Instance 3($\beta=0.3772$, $p=0.0063$). Similarly, at Instance 2, better performance was shown for better paired associate learning ($\beta=0.3403$, $p<0.0001$ ), TMT-B ($\beta=-16.716$, $p=0.014$), and Reaction time ($\beta=-5.426$, $p=0.0186$ ), consistent with enhanced memory, learning, and executive function. Furthermore, cognitive domains such as prospective memory and numeric memory showed weaker but positive associations which supports cognitive benefits of engaging in MVPA at or above recommended levels.

In fully adjusted models, higher MVPA (minutes/week) was associated with better cognitive performance, particularly for fluid intelligence ($\beta$=0.0030) and paired associate learning ($\beta$=0.0031) at Instance 3 (Figure~\ref{fig:mvpa_cognition_combined}, Table S.4). Similar association were shown using categorical (WHO-based) MVPA, where guideline adherence was associated with positive—though nonsignificant—associations (Table S.4) The consistent direction of associations suggests a potential link between MVPA and cognitive health withshared influences by other factors like education, socioeconomic status, and health behaviors.

\subsection{Brain Regions}
\subsubsection{Subcortical Volumes}
\begin{figure}[htbp]
    \centering
    \begin{subfigure}[t]{\textwidth}
        \centering
        \hspace*{-1.5cm}
        \includegraphics[width=1\linewidth]{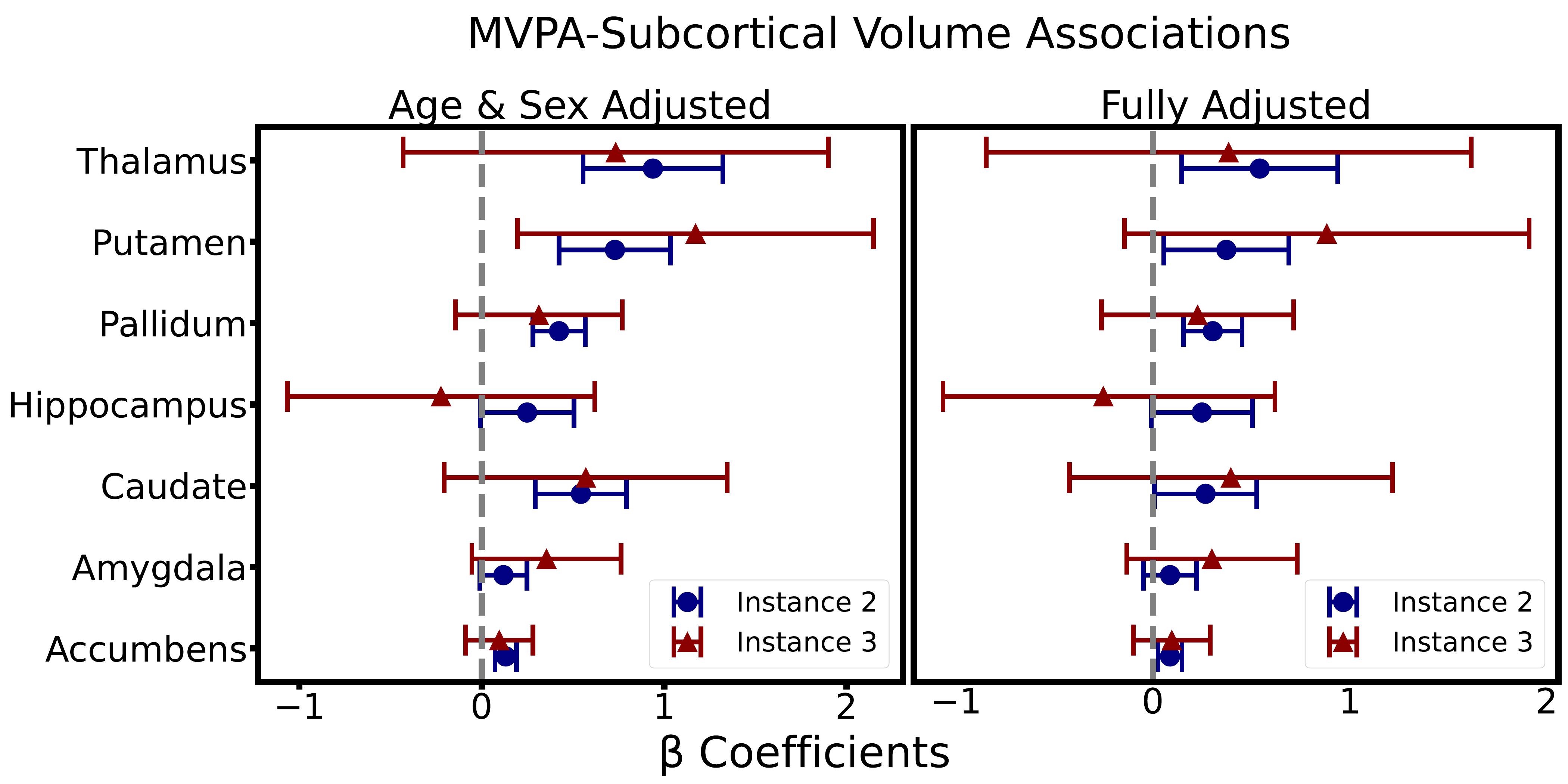}
        \label{fig:subfig1}
    \end{subfigure}
   \vspace{0.2cm}  
    \begin{subfigure}[t]{\textwidth}
        \centering
        \includegraphics[width=0.85\linewidth, trim=10 990 20 20, clip]{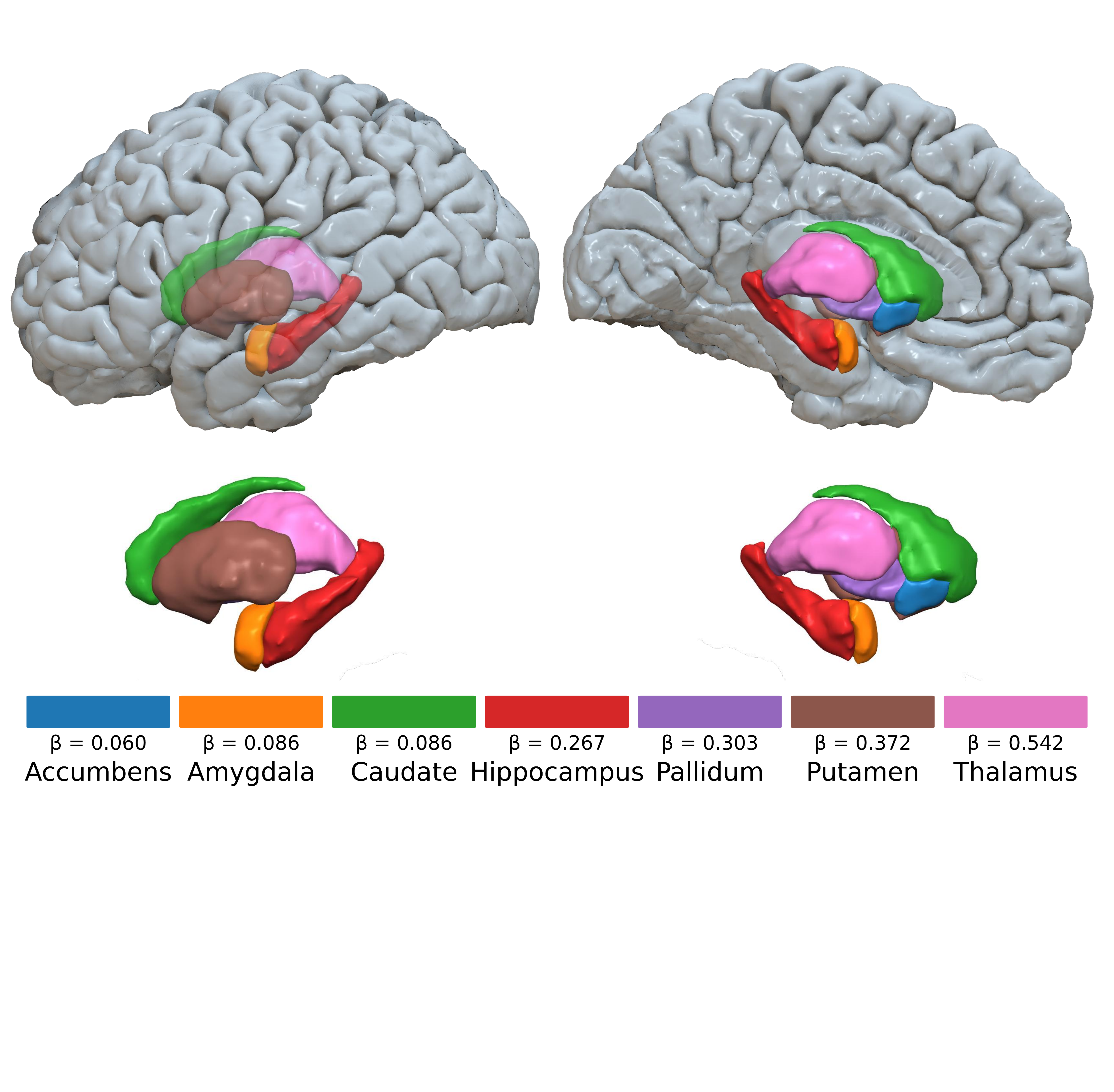}
        \label{fig:subfig2}
    \end{subfigure}
   \caption{Associations between physical activity and brain region volumes for base and fully adjusted models. Lower part of the figure presents subcortical volume associations using fully adjusted model rendered using the Surfice tool, with each region distinctly colored for visual identification. Covariates for fully adjusted models include age (linear and quadratic), sex, ethnicity, smoking status, alcohol use, Townsend deprivation index, diabetes history, HDL and LDL cholesterol, education (qualifications), systolic and diastolic blood pressure, triglycerides, longstanding illness, and BMI.}
    \label{fig:brain_mvpa}
\end{figure}

In base model, higher MVPA was positively associated with volumes across several subcortical regions (Figure~\ref{fig:brain_mvpa}, Table~S.5). At Instance~2, MVPA was associated with caudate ($\beta=0.5431$, $p<0.0001$), putamen ($\beta=0.7297$, $p<0.0001$), thalamus ($\beta=0.9380$, $p<0.0001$), and pallidum ($\beta=0.4232$, $p<0.0001$), nucleus accumbens ($\beta=0.1311$, $p<0.0001$), amygdala ($\beta=0.1182$), and hippocampus ($\beta=0.2481$). At Instance~3, associations were directionally consistent with some non-significant associations likely due to reduced sample size ($n=745$) and greater variance (Figure~1, Table~1).
WHO-based MVPA categories yielded similar results at Instance~2, with higher volumes in the thalamus ($\beta=58.60$, $p<0.0001$), putamen ($\beta=50.23$, $p<0.0001$), caudate ($\beta=39.96$, $p<0.0001$), pallidum ($\beta=30.53$, $p<0.0001$), amygdala ($\beta=11.91$, $p=0.0173$), hippocampus ($\beta=20.61$, $p=0.375$), and nucleus accumbens ($\beta=8.08$, $p<0.001$).
In fully adjusted models (Figure~\ref{fig:brain_mvpa}, Table S.6), MVPA remained positively associated with several subcortical volumes such as nucleus accumbens ($\beta=0.0861$, $p=0.0057$), caudate ($\beta=0.2667$, $p=0.0435$), pallidum ($\beta=0.3032$, $p=0.0001$), putamen ($\beta=0.3720$, $p=0.0216$), and thalamus ($\beta=0.5419$, $p=0.0073$) at Instance~2. For categorical MVPA, positive association were shown for caudate ($\beta=19.88$, $p=0.0457$), pallidum ($\beta=21.00$, $p=0.0003$), and amygdala at Instance~3 ($\beta=35.46$, $p=0.0372$). Other subcortical brain regions also show positive association with MVPA which support the potential relationship between MVPA and subcortical brain volumes. To explore lateralization effects, we performed experiment to examine left and right hemispheres separately (Tables S7–S10). The results show consistent and positive associations with several asymmetries and stronger association in the right accumbens and thalamus, and left caudate and pallidum (more details in Appendix pp 3-4).

\subsection{Regional Grey Matter Volumes}

MVPA (mins/week and based WHO guidelines) show positive associations across multiple regional gray matter volumes for both based and fully adjusted model (Appendix p.4-5, Table A1-A4). Multile regional gray matter volumes was positively associated, for instance, in fully adjusted models:  medial frontal cortex (left: $\beta=0.263$, $p=0.008$; right: $\beta=0.281$, $p=0.005$), frontal pole (left: $\beta=1.645$, $p=0.034$; right: $\beta=1.950$, $p=0.024$), insular cortex (left: $\beta=0.463$, $p=0.021$; right: $\beta=0.435$, $p=0.031$), angular gyrus (right: $\beta=0.786$, $p=0.026$), and Crus I of the right cerebellum ($\beta=1.321$, $p=0.012$) and parahippocampal gyrus (left: $\beta=0.495$, $p<0.001$; right: $\beta=0.471$, $p<0.001$). 
These regions 
are responsible for emotion processing, working memory, and high-level visual perception, highlighting the importance of preserving gray matter in areas vulnerable to aging and neurodegeneration~\cite{burns2008cardiorespiratory,takeuchi2018impact,won2022differential,nyatega2022gray}. Detailed results are represented in Appendix (p.4-5, Table A1-A4).

\section{Sensitivity and Subgroup Analysis}
We categorized participants MVPA duration to to assess robustness of associations between MVPA and subcortical brain volumes and potential dose–response effects (Table S.11). In fully adjusted models, higher MVPA were consistently associated with larger subcortical volumes, particularly at Instance 2. Participants engaging in $>$300 minutes/week had significantly greater volumes in the nucleus accumbens ($\beta=11.20$, $p=0.0045$), pallidum ($\beta=27.27$, $p=0.0047$), putamen ($\beta=59.02$, $p=0.0040$), and thalamus ($\beta=63.47$, $p=0.0131$), suggesting a dose–response pattern. At Instance 3, associations remained directionally consistent, with significant effects for the nucleus accumbens and putamen, though power was limited by a smaller sample size (N=745 vs. N=8387 at Instance 2). These findings may suggest the neuroprotective link between MVPA and subcortical structures and highlight the robustness of these effects. 
To further examine the robustness and generalizability of the associations between MVPA, cognitive function, and brain structure, we conducted stratified analyses across key subgroups, adjusting for all relevant confounders except the stratifying variable in each model. Specifically, for cognitive functions, we analyzed participants under and above the age of 65 (Table S12-13), and sex (Tables S14–S15) and obesity status (Tables S16–S17). Similarly, analysis was performed for subcortical brain regions (Table S.18-Table S.23). For instance, In adults under 65 (Table S.18), MVPA was positively associated with volumes in nearly all subcortical regions at instance 2. Among those aged 65 and over (Table S.19), associations were also  positive including nucleus accumbens, pallidum, and thalamus. Sex-stratified models showed stronger associations in men (such as accumbens, pallidum, and putamen), while in women, positive effects were seen for the amygdala, pallidum, and thalamus. Associations were generally stronger in non-obese individuals, though MVPA was associated with greater caudate volume in the obese group.

\subsection{Physical Activity timing}
\begin{figure}
    \centering
    \includegraphics[width=1\linewidth]{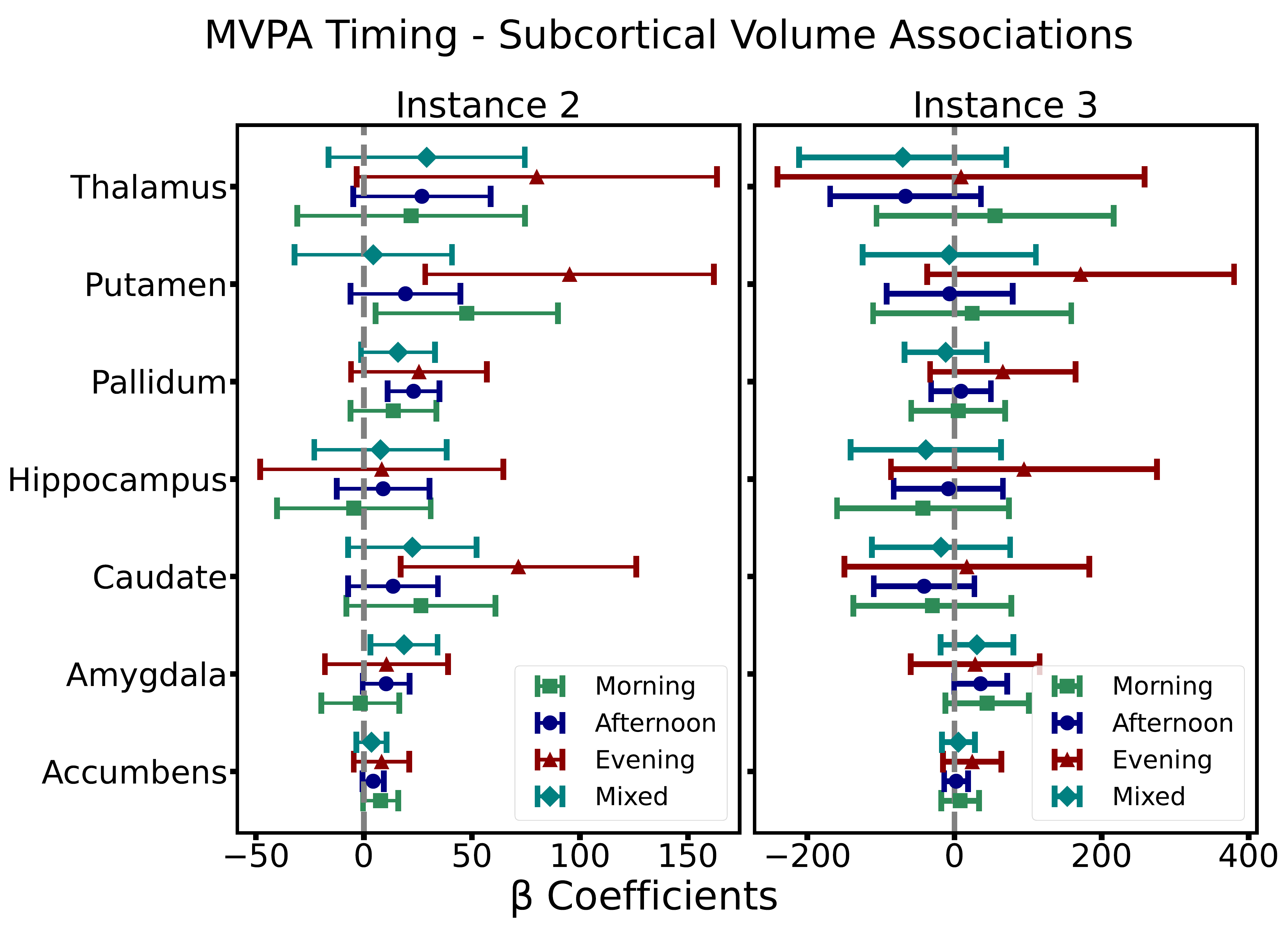}
    \caption{Associations between timing of physical activity and average brain region volumes. Reference group: inactive individuals. Results are based on fully adjusted models controlling for age (linear and quadratic), sex, ethnicity, smoking, alcohol use, Townsend deprivation index, diabetes history, HDL and LDL cholesterol, education, blood pressure (systolic and diastolic), triglycerides, longstanding illness, and BMI.}
    \label{fig:mvpatiming}
\end{figure}

For secondary analyses, we utilized fully adjusted models which show MVPA was associated with cognitive function and brain regions at all times of the day compared to inactive individuals. For instance,  for cognitive function (Table A.5), midday afternoon and evening MVPA were associated with paired associate learning at Instance 2 ($\beta$=0.1325, p=0.048) and Instance 3 ($\beta$=0.4630, p=0.019). MVPA was also associated with subcortical regions such as at Instance 2 (Figure\ref{fig:mvpatiming}, Table A.6), larger caudate and putamen with evening MVPA, and greater pallidum and amygdala volume with afternoon or mixed activity. All MVPA timing groups showed greater grey matter volumes than inactive individuals, particularly in frontal, temporal, cingulate, insular cortices, parahippocampal gyrus, and amygdala (Table A.7, all $p<0.05$), suggesting that MVPA—regardless of timing—is associated with preserved brain structure (details in Appendix pp.5-6, Tables A5–A7).

\section{Discussion}
This study based on UKB data with over 45,000 older adults found consistent associations of higher MVPA with better cognitive performance, subcortical, and regional gray matter volumes. Our findings show dose–response and timing-specific effects of MVPA on both cognitive function and brain structure. Moreover, our results show region-specific and lateralized effects associated with MVPA, offering new insight into how PA may influence cognitive aging.

Our findings current guidelines recommending at least 150 minutes of MVPA per week to promote brain health in later life~\cite{livingston2020dementia,world2024nearly, strain2024national}. Our results show that higher MVPA was associated with better cognitive performance such as faster reaction time, fluid intelligence, Trail Making Tests, and paired associate learning. These associations were statistically significant in the base model, and while attenuated in fully adjusted models, the overall directionality remained consistent which suggests that MVPA is beneficial for cognitive health through mechanisms beyond traditional socioeconomic and health-related factors. 

Neuroimaging analyses show that higher MVPA was associated with several brain regions, particularly subcortical structures and frontal-temporal cortical areas. Positive association was persistent in base and fully adjusted models in multiple subcortical brain regions such as caudate, putamen, thalamus, and pallidum which are critical for motor coordination, cognitive control, and motivation. Although most associations was bilateral, we observed asymmetries between the hemispheres (nucleus accumbens and caudate). MVPA was also associated with grey matter volumes in 124 regions with positive associations, including 38 regions with significant associations in instance 2. These regions includes prefrontal, temporal, Cerebellum, including the medial frontal cortex, and parahippocampal gyrus which are cirtical for executive function, memory, emotion regulation, and cognitive flexibility. These medial temporal regions are among the earliest affected by aging and Alzheimer’s disease, suggesting that MVPA may buffer against age-related atrophy~\cite{frisoni2010clinical}. Furthermore, gray matter is more vulnerable to Alzheimer’s disease, underscoring the importance of maintaining gray matter integrity~\cite{burns2008cardiorespiratory}. In addition, the insula supports emotion processing~\cite{takeuchi2018impact}, cerebellum—particularly lobules VI, Crus I, and VIIIa—is associated with working memory in structural and functional imaging studies~\cite{won2022differential}. Furthermore, the fusiform gyrus, known for high-level visual processing, shows reduced gray matter in individuals with cognitive impairment and Parkinson’s disease, and associated with poor visuoperceptual performance~\cite{nyatega2022gray}. These findings reinforce the relevance of MVPA in preserving gray matter in regions supporting cognitive, emotional, and perceptual functions vulnerable to aging and neurodegenerative decline. These results suggest that MVPA promotes global brain health: larger cortical and subcortical structures, and intact white matter pathways in physically active older adults~\cite{hamer2018association, melo2023objective, hofman2023physical}.

MVPA at any time of the day (morning, midday-afternoon, or evening) was beneficial, suggesting the flexibility of when MVPA can be performed to support brain health. Subgroup and sensitivity analyses further support the generalizability of our findings. Associations of MVPA with cognition and subcortical volumes were observed across age, sex, and obesity, although effect sizes were slightly larger among younger and non-obese individuals. Our dose–response analysis demonstrates that even modest MVPA levels—below current WHO guidelines—are associated with measurable cognitive and brain structural benefits, suggesting that partial adherence may still offer neuroprotective value. However, greater benefits were observed at higher MVPA levels, particularly above 300 minutes per week. This highlights both the accessibility of initial gains through modest activity and the additional value of exceeding current guidelines for optimal brain health.

Consistent with prior work, our results suggest that regular MVPA may protect against cognitive decline and dementia, as higher PA levels are associated with greater hippocampal volumes and reduced dementia risk~\cite{hofman2023physical,hamer2018association, sofi2011physical,zhu2017objectively}. Meta-analyses estimate that higher activity levels are associated with a 15–20\% reduction in all-cause dementia risk~\cite{iso2024physical,iso2022physical}. However, prior studies either used self reported PA, lacked MVPA measures, did not utilize cognitive data~\cite{hofman2023physical,hamer2018association, sofi2011physical,zhu2017objectively}, or estimates are limited by heterogeneity in PA measurement across studies~\cite{iso2024physical,iso2022physical}. Moreover, unlike studies that model dementia incidence over follow-up, our approach captures variation in cognitive and brain health across community-dwelling older adults, without restricting the sample to those who will later develop dementia. It aid in a broader understanding of how PA may influence cognitive and brain aging in older adults. These findings underscore the importance of objectively measured MVPA in understanding its role in promoting cognitive reserve and resilience to neurodegeneration. 


PA supports not only better health but also greater productivity, with implications for economic performance on both individual and societal levels. Higher PA levels are consistently linked to better cognitive outcomes—for example, meta‐analyses show that highly active adults have approximately 38\% lower risk of cognitive decline (Hazard  Ratio$\approx 0.62$, 95\% CI 0.54–0.70)~\cite{sofi2011physical} and approximately 14\% lower incidence of dementia (Relative Risk$\approx 0.86$)~\cite{blondell2014does}. Such cognitive improvements have been linked in prior studies to better workforce outcomes: for instance a one‐standard‐deviation increase in cognitive test scores is associated with wage increases of approximately 4.5\% (95\% CI 2.6–9.6\%)~\cite{ozawa2022educational} and an increase in the likelihood of white‐collar employment~\cite{ozawa2022educational}. In contrast, poor cognitive function (to which inactivity contributes) is associated with lower job retention and weaker decision‐making and financial management skills. At a macroeconomic level, increasing population PA could substantially boost productivity \cite{sofi2011physical,hafner2019economic}. RAND modeling estimates that improvements in physical activity could generate global GDP gains of USD 138–338 billion by 2025 and USD 314–760 billion by 2050, with cumulative gains reaching up to USD 14.4 trillion by midcentury~\cite{hafner2019economic}. These projections are largely driven by reductions in absenteeism and presenteeism, as physically active individuals consistently report fewer missed workdays and less on-the-job productivity loss~\cite{walker2017longitudinal}. PA also eases healthcare burdens: global healthcare savings from activity-related disease prevention are estimated at USD 8.7–11.2 billion annually, rising to USD 16–20.6 billion by 2050~\cite{hafner2019economic}. In contrast, physical inactivity is linked to significant burdens for individuals and health systems alike . For instance, inactive Finnish adults incurred \EUR{4,300} more per year in combined health and productivity costs than active peers~\cite{t2022individual}, and globally, inactivity contributes to 13–15 million  disability-adjusted life-years (DALYs) annually~\cite{ding2016economic}. Together, these findings position PA as both a public health priority and a potential driver of long-term economic benefits, with possible downstream impacts on cognitive function, workplace productivity, and population health.

Our findings have important clinical and public health implications for promoting healthy cognitive aging. As dementia is incurable, delaying it even by small magnitude can yield substantial population-level benefits. Thus encouraging MVPA in mid- and late-life is a low cost and scalable strategy that may reduce dementia burden by improving brain network efficiency and cognitive reserve~\cite{konwar2023effect}. Additionally, active lifestyles are often associated with other factors that support brain health, such as social engagement, better mood, and improved sleep, which further boost cognitive function. Our results emphasize the importance of exercise counseling in geriatric care, advising patients that adhering to WHO activity guidelines may help preserve memory and daily functioning. At the policy level, these findings support the integration of PA promotion into dementia prevention and healthy aging initiatives. Given that over 30\% of adults especially 60 years and older globally do not meet activity guidelines \cite{strain2024national}, there is need for improvement. Urban planning, community programs, and healthcare systems should promote safe exercise opportunities particularly for seniors to address this modfiable risk factor including smoking, alcohol consumption and vascular health. However, given the observational design, causality cannot be confirmed, and reverse causation—where early cognitive decline reduces activity—remains a possibility for future investigation.

Key strengths of this study include device-based objective MVPA, comprehensive cognitive and neuroimaging assessments, and robust confounder adjustment (e.g., age, sex, education). Sensitivity, and subgroup analysis including MVPA timing further enhances generalizability, and the alignment between cognitive and neuroimaging findings supports the true MVPA effect. However, limitations include the cross-sectional design, which precludes causal inference. Although we adjusted for multiple covariates, however, residual confounding still remains possible such as early cognitive decline may reduce activity (reverse causation), or shared genetic/socioeconomic status factors may influence both activity and brain health. Additionally, accelerometer data was only collected for seven days which reflects short-term behavior and may not capture lifetime activity patterns. However, the consistent, statistically significant associations, and established research suggest MVPA is a meaningful contributor to cognitive aging. Moreover, longitudinal analysis was not performed due to uneven follow-up and reduced statistical power from smaller sample sizes at later visits.Future longitudinal and interventional studies should include long-term cohorts and randomized exercise trials incorporating cognitive, neuroimaging, and mechanistic biomarkers (e.g., BDNF, PET, cerebrovascular measures). Trials stratified by genetic risk (e.g., APOE), or with more robust exclusion criteria (e.g., pre-existing conditions like dementia or Parkinson’s disease) may help identify subgroups most likely to benefit. Ultimately, demonstrating that exercise slows cognitive decline would support embedding MVPA in cognitive health guidelines. Regular MVPA should be promoted as a low-risk, high-reward dementia prevention strategy—an approach supported by our findings and broader literature.

\section{Conclusion}
Higher levels of MVPA were associated with improved cognitive performance and preservation of brain structure in older adults. These findings highlight MVPA as a modifiable factor that may delay cognitive decline, support functional independence, and reduce dementia burden. Encouraging regular activity in mid- and late-life should be prioritized in clinical care and public health policy as part of dementia prevention and healthy aging strategies.






\section*{Competing Interests}
The authors declare that the research was conducted in the absence of any commercial or financial relationships that could be construed as a potential conflict of interest.


\section*{Acknowledgements}
This work is supported by the National Science Foundation under Grant No.2123809 and National Institutes of Health/National Institute on Aging (RF1AG071469, R01AG071469). The content is solely the responsibility of the authors and does not necessarily represent the official views of the National Institutes of Health and the National Science Foundation.


\section*{Data availability statement}
The data that support the findings of this study are available from the UK Biobank, but restrictions apply to the availability of these data. We used the UK Biobank dataset under the approved application ID  48388. The data is not publicly available but can be accessed from the UK Biobank team upon application approval.


\bibliographystyle{unsrt}
\bibliography{sample}



\section*{Supplementary material}

# Supplementary Materials

## Contents

## Tables

# Figures

# Dataset Description (Methods)

The UK Biobank (UKB) is a large-scale, population-based prospective cohort study that enrolled 500,000 participants aged 40 to 69 at baseline, recruited between 2006 and 2010 across 22 assessment centers in the UK [1]. The UKB

collects extensive information on sociodemographic characteristics, lifestyle behaviors, health status, and physical measurements, with data linkage to medical records [2].

Between 2013 and 2015, a subset of 103,612 participants consented to participate in an accelerometer sub-study, where they were asked to wear a wrist-worn triaxial accelerometer (Axivity AX3) continuously for seven days to objectively measure PA [3]. The device captured raw acceleration data at a sampling frequency of 100 Hz and a dynamic range of ±8g. To ensure reliable recordings, devices were pre-programmed to initiate data collection at 10 a.m. two weekdays following postal dispatch, minimizing the chance of collecting data during transit [3]. Participants returned devices by mail after the monitoring period. Details of the recruitment, data acquisition protocols, and device calibration have been described in detail in prior publications. Data from the accelerometers were processed by the UKB accelerometer team to derive PA metric. MVPA was identified using a machine learning classification model trained to distinguish MVPA from other activity states (e.g., sedentary, light activity, sleep). The validity of this approach has been previously demonstrated in UK-based populations, with reported accuracy exceeding 85% [4].

Our study utilized data from the UKB accelerometer sub-study, conducted under approved UKB Application 48388. MVPA exposure in the current study was assessed using continuously (mins/week) and categorically (thresholded using ≥150 min/week based on WHO guidelines). MVPA was computed during waking hours, defined as 5:00 a.m. to midnight, to minimize inclusion of movement during typical sleep periods [5,6]. To ensure the reliability of activity estimates, participants who did not provide a full week of data had their MVPA values extrapolated based on their average daily activity. We employed inclusion and exclusion criteria. Participants were excluded if they (1) had withdrawn consent from the UK Biobank, (2) lacked activity data for any hour in the 24-hour cycle, or (3) exhibited excessive nocturnal activity (defined as >10% of daily MVPA between 01:00 and 04:00) [5,6]. In addition, individuals were excluded for having poor-quality accelerometer data, which included unreliable file size (i.e., abnormally small or large; Field ID: 90002); wear time of less than 72 hours or failure to cover all 24-hour periods over the 7-day span (Field ID: 90015); uncalibrated or poorly calibrated devices (Field ID: 90016); devices reused without proper recalibration (Field ID: 90017); data with recording interruptions (Field ID: 90180); or those with excessive data errors, defined as >768 anomalies based on interquartile range thresholds (Field ID: 90182). Additionally, participants with missing data on any of the covariates used in the models were excluded to ensure complete-case analysis.

# Results

## Subcortical Regions

In the base model, higher MVPA were positively associated with volumes across several subcortical brain regions (Table S.5) such as caudate (Instance 2: $\beta = 0.5431$, $p < 0.0001$), putamen (Instance 2: $\beta = 0.7297$, $p < 0.0001$; Instance 3: $\beta = 1.1717$, $p = 0.0186$), thalamus (Instance 2: $\beta = 0.9380$, $p < 0.0001$), and pallidum (Instance 2: $\beta = 0.4232$, $p < 0.0001$), nucleus accumbens (Instance 2: $\beta = 0.1311$, $p < 0.0001$), amygdala (Instance 2: $\beta = 0.1182$), and hippocampus (Instance 2: $\beta = 0.2481$), suggesting that greater MVPA may correspond with increased subcortical volume in these regions. Effects were generally smaller and less consistent at instance 3, likely reflecting lower sample sizes (n=745 at instance 3) and greater variance. When MVPA was categorized according to WHO guidelines, individuals meeting recommended activity levels had higher subcortical volumes across several regions at Instance 2: thalamus ($\beta = 58.60$, $p < 0.0001$), putamen ($\beta = 50.23$, $p < 0.0001$), caudate ($\beta = 39.96$, $p < 0.0001$), pallidum ($\beta = 30.53$, $p < 0.0001$), amygdala ($\beta = 11.91$, $p = 0.0173$), hippocampus ($\beta = 20.61$, $p = 0.375$), and nucleus accumbens ($\beta = 8.08$, $p < 0.001$). These associations were consistent with those observed in continuous MVPA models, reinforcing the relationship between PA and subcortical brain structure. At Instance 3, although most estimates remained directionally positive, however, non-significant—likely driven by reduced sample size and increased noise. In fully adjusted models (Table S.6), positive associations were observed between higher MVPA and several subcortical brain regions. When MVPA was modeled as a continuous variable, greater activity was associated with larger average volumes of the nucleus accumbens ($\beta = 0.0861$; $p = 0.0057$), caudate ($\beta = 0.2667$; $p = 0.0435$), pallidum ($\beta = 0.3032$; $p = 0.0001$), putamen ($\beta = 0.3720$; $p = 0.0216$), and thalamus ($\beta = 0.5419$; $p = 0.0073$) at Instance 2, even after adjustment. These results indicate that daily MVPA may contribute to better structural brain integrity, particularly in basal ganglia and thalamic regions. Using WHO guideline-based MVPA categories, individuals meeting recommended activity levels showed

larger subcortical volumes at Instance 2, including the caudate (β = 19.88, p = 0.0457), pallidum (β = 21.00, p = 0.0003), and amygdala at Instance 3 (β = 35.46, p = 0.0372), suggesting that higher MVPA may relate to structural brain differences with potential neuroprotective relevance. While associations for other regions (e.g., hippocampus, nucleus accumbens, putamen, and thalamus) did not reach statistical significance in fully adjusted models, effect directions were consistently positive and aligned with base model results. This pattern supports a possible underlying relationship between MVPA and brain structure, though associations may be influenced by shared variance with factors such as education, socioeconomic status, and comorbidities. To explore lateralization effects, we examined left and right hemispheres separately across modeling strategies (Tables S7–S10). Most associations were bilateral and consistent across models, though some asymmetries emerged—stronger effects in the right accumbens and thalamus, and left caudate and pallidum—particularly in base models, with partial attenuation in fully adjusted model.

## Regional Grey Matter Volumes

We assessed the association between MVPA, and regional grey matter volumes derived from T1-weighted structural MRI using FAST, using base and-fully adjusted model. Continuous MVPA measures were positively associated with grey matter volume in several brain regions, particularly those involved in cognitive, emotional, and motor functions. Notably, higher MVPA levels were associated with increased grey matter volume in the amygdala, hippocampus, and prefrontal cortex [7,8]. MVPA was positively associated with volumes in multiple regions (Table A1), including the right amygdala (β = 0.387, p < 0.00001) and hippocampus (β = 0.313, p = 0.0128), suggesting preserved integrity in areas linked to emotion and memory. In the prefrontal cortex, positive associations were seen in the left and right medial frontal cortex (left: β = 0.328, p = 0.00061; right: β = 0.383, p < 0.0001), and right frontal pole (β = 2.899, p < 0.0006), supporting links to executive function and planning. Strong associations were also observed in cerebellar Crus I and II (all p < 0.0001). Additional effects in the insula, precuneus, temporal cortex, and ventral striatum suggest widespread MVPA-related structural benefits. A negative association in the brainstem (β = −0.476, p = 0.0357) may reflect regional sensitivity or confounding.

In the fully adjusted model (Table. A3), several associations remained significant, suggests the robustness of the relationship between MVPA and brain structure. Higher MVPA was positively associated with grey matter volume in the medial frontal cortex (left: β = 0.263, p = 0.008; right: β = 0.281, p = 0.005) and frontal pole (left: β = 1.645, p = 0.034; right: β = 1.950, p = 0.024), consistent with findings with the based model. These regions are critical for executive control, planning, and emotion regulation [9-11]. Additional significant associations were shown in the insular cortex (left: β = 0.463, p = 0.021; right: β = 0.435, p = 0.031), angular gyrus (right: β = 0.786, p = 0.026), and Crus I of the right cerebellum (β = 1.321, p = 0.012), supporting the hypothesis that MVPA confers widespread neuroprotective effects. Further effects were found in temporal regions including the middle temporal gyrus (left: β = 0.447, p < 0.001; right: β = 0.288, p = 0.003) and parahippocampal gyrus (left: β = 0.495, p < 0.001; right: β = 0.471, p < 0.001). Associations based on WHO-defined MVPA categories showed a similar pattern to the continuous MVPA model, with higher activity levels consistently associated with more favorable cognitive and brain outcomes (see Table. A2, Table. A4). Participants meeting WHO PA recommendations showed higher grey matter volumes in several regions, including the hippocampus and amygdala, consistent with effects observed in age- and sex-adjusted models. These structures are central to memory and emotional regulation. Additional volumetric preservation was observed in the frontal cortex, cingulate gyrus, cerebellum, insula, and precuneus—regions important for cognitive control, emotional processing, and self-awareness [12,13]. In the fully adjusted model, the observed positive relationships between regional grey matter volumes and MVPA remained consistent (see Table A2, Table A4). These results support previous literature [7,8] and reinforce the idea that MVPA contributes to widespread neuroprotective effects, likely benefiting cognition, mood, and motor function.

## Physical Activity timing

In a secondary analysis examining the timing of MVPA, for fully adjusted models, MVPA timing showed generally positive—though largely nonsignificant—associations with cognitive performance, particularly at instance 0 (Table

A.5). Compared to inactive individuals, those with middat-afternoon or Evening MVPA profiles tended to exhibit better cognitive function across several domains. At Instance 2, Afternoon MVPA was positively associated with paired associate learning ($\beta = 0.1325$, p = 0.048), with a stronger effect at Instance 3 ($\beta = 0.4630$, p = 0.019). For other cognitive domains, consistent positive associations were observed across MVPA timing profiles. We also observed time- and region-specific associations between MVPA timing and subcortical brain volumes in Instance 2 (Table A.6). Evening MVPA was significantly associated with caudate ($\beta = 71.49$, p=0.0102) and putamen volumes ($\beta = 95.21$, p=0.0052), while midday-afternoon MVPA showed robust associations with greater pallidum ($\beta = 22.99$, p=0.0002) and trend-level associations with amygdala volume ($\beta = 10.30$, p=0.063). Mixed MVPA was linked to increased amygdala volume ($\beta = 18.58$, p=0.019). These findings suggest that MVPA performed later in the day, particularly in the afternoon or evening, may confer structural benefits to subcortical regions that support cognitive health. Furthermore, compared to inactive individuals, all MVPA timing groups show significantly larger grey matter volumes in multiple brain regions (Table A.7), including the frontal, temporal, cingulate, and insular cortices, as well as the parahippocampal gyrus and amygdala (p <0.05). These findings suggest that engaging in MVPA, regardless of timing, is associated with greater brain volume in regions critical for cognition and emotion.

*Table S. 1 Detailed Description and Interpretation of Cognitive Performance Variables from the UK Biobank. All cognitive tests were administered on supervised kiosks at assessment centers, typically following health questionnaires and preceding interviews. The Snap test, a reaction time assessment, measured simple processing speed by recording the average time (in milliseconds) taken to press the button during trials with correct paired matches. The fluid intelligence test comprised 13 language and mathematics questions, with the raw score calculated as the unweighted sum of correct responses. The prospective memory test evaluated memory and inhibitory control by instructing participants to perform a specific action later in the assessment; performance was dichotomized as either correct on the first attempt or incorrect. Additional cognitive assessments included numeric memory, trail making tests A and B (TMTA and TMTB), and paired associate learning, each evaluating different cognitive domains such as working memory, executive function, processing speed, and associative memory.*

| Cognitive Variable | Cognitive Domain | Description | Positive Estimate Interpretation | Negative Estimate Interpretation | Preferred Direction |
|---|---|---|---|---|---|
| Reaction Time | Processing Speed | Participants completed a timed test of symbol matching, similar to the card game 'Snap'. Faster completion times indicate better processing speed. | Slower reaction (worse processing speed) | Faster reaction (better processing speed) | Negative (↓ Lower is better) |
| Fluid Intelligence | Reasoning | A task with thirteen logic/reasoning-type questions and a two-minute time limit, assessing verbal and numerical reasoning abilities. | Indicates higher reasoning abilities (better performance). | Indicates lower reasoning abilities (worse performance). | Positive (↑ Higher is better) |
| Numeric Trail Duration | Executive Function | Participants completed timed tasks involving numeric sequences. Faster completion times indicate better executive function. | Indicates faster task completion (better | Indicates slower task completion (worse performance) | Negative (↓ Lower is better) |
| Alphanumeric Trail Duration | Executive Function | Participants completed timed tasks involving alphanumeric sequences. Faster completion times indicate better executive function. | | | |
| Paired associate learning | Verbal Memory | Participants were tasked with associating word pairs. It is the number of word pairs correctly associated out of ten attempts. | Indicates better verbal memory (higher score) | Indicates poorer verbal memory (lower score). | Positive (↑ Higher is better) |
| Maximum Digits Recalled | Working/numeric Memory | Participants were shown a sequence of digits and asked to recall them in the correct order. The length of the sequence increased until the participant made an error. | Indicates better working memory capacity (higher score). | Indicates lower working memory capacity (lower score) | Positive (↑ Higher is better) |

*Table S. 2. Details of the variables, exposures, covariates, and data sources used in this study.*

| Variable | Field ID | Notes |
| --- | --- | --- |
| Sex | 31 | acquired from central registry at recruitment |
| Ethnicity | 21000 | |
| Townsend deprivation index | 22189 | at recruitment |
| Alcohol intake frequency | 1558 | |
| Wear duration overall | 90051 | |
| Year of birth | 34 | Used to calculate age at accelerometry |
| Month of birth | 52 | |
| Start time of wear | 90010 | |
| UK Biobank assessment Center | 54 | |
| Long-standing illness, disability or infirmity | 2188 | |
| Diabetes diagnosed by doctor | 2443 | |
| Cancer diagnosed by doctor | 2453 | |
| Body mass index (BMI) | 21001 | |
| Smoking status | 20116 | |
| Derived accelerometry | 1020 | |
| Qualifications | 6138 | |
| HDL cholesterol | 23406 | |
| LDL direct | 30780 | |
| Diastolic blood pressure | 4079 | |
| Systolic blood pressure | 4080 | |
| Triglycerides | 30870 | |
| Regional grey matter volumes (FAST) | 1101 | |
| Subcortical volumes (FIRST) | 1102 | |
| Reaction time | 20023 | Mean time to correctly identify matches |
| Fluid intelligence | 20016 | Fluid intelligence score |
| Trail making Test A | 6348 | Duration to complete numeric path (trail #1) |
| Trail making Test B | 6350 | Duration to complete alphanumeric path (trail #2) |
| Prospective memory result | 20018 | |
| Paired associate learning | 20197 | Number of word pairs correctly associated |
| Numeric memory | 4282 | Maximum digits remembered correctly |

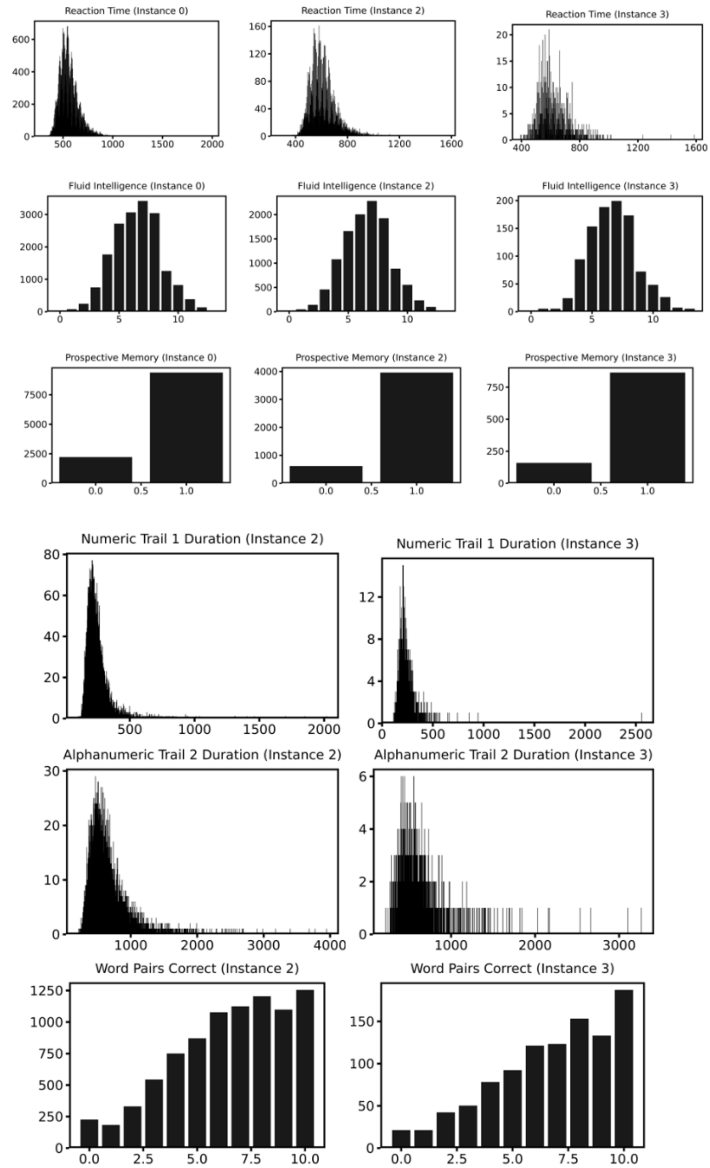

*Figure S. 1. Histogram of cognitive variables from the UK Biobank.*

*Table S. 3. Associations between MVPA and cognitive function (adjusted for age and sex). Bold results show significant (p<0.05) results.*

| Cognitive Variables | MVPA in mins per week | | WHO Guidelines | | N_Obs |
|---|---|---|---|---|---|
| | Estimate (95% CI) | P_Value | Estimate (95% CI) | P-value | |
| Reaction Time | | | | | |
| Instance 0 | -0.0160 (-0.0456 to 0.0137) | 0.2922 | -1.8652 (-3.9568 to 0.2265) | 0.0805 | 45784 |
| Instance 2 | 0.0002 (-0.0587 to 0.0590) | 0.9959 | -5.4260 (-9.9443 to -0.9077) | 0.0186 | 11520 |
| Instance 3 | 0.1508 (-0.0412 to 0.3428) | 0.1236 | -6.3588 (-21.5859 to 8.8683) | 0.4127 | 1002 |
| **Fluid Intelligence Instance 0** | **0.0031 (0.0022 to 0.0040)** | **<0.0001** | **0.2454 (0.1823 to 0.3086)** | **<0.0001** | 17651 |
| **Fluid Intelligence Instance 2** | **0.0027 (0.0017 to 0.0038)** | **<0.0001** | **0.1909 (0.1095 to 0.2724)** | **<0.0001** | 11351 |
| **Fluid Intelligence Instance 3** | **0.0052 (0.0017 to 0.0086)** | **0.0031** | **0.3772 (0.1069 to 0.6476)** | **0.0063** | 1001 |

| | | | | | |
|---|---|---|---|---|---|
| Trail making test- A | | | | | |
| **Instance 2** | **-0.0589 (-0.1145 to -0.0034)** | **0.0376** | -0.1086 (-4.4455 to 4.2283) | 0.9608 | 8546 |
| Instance 3 | 0.0967 (-0.0879 to 0.2814) | 0.3042 | -0.8380 (-15.4439 to 13.7680) | 0.9104 | 998 |
| Trail making test- B | | | | | |
| Instance 2 | -0.1486 (-0.3181 to 0.0210) | 0.0859 | **-16.7160 (-30.0144 to -3.4186)** | **0.0138** | **8302** |
| Instance 3 | -0.0996 (-0.5946 to 0.3954) | 0.6930 | -14.0010 (-53.2940 to 25.2910) | 0.4845 | 976 |
| **Paired associate learning Instance 2** | **0.0027 (0.0012 to 0.0043)** | **0.0006** | **0.3403 (0.2185 to 0.4622)** | **<0.0001** | 8646 |
| **Paired associate learning Instance 3** | **0.0046 (0.0002 to 0.0090)** | **0.0397** | **0.5277 (0.1794 to 0.8761)** | **0.0030** | 1021 |
| **Numeric memory Instance 0** | 0.0009 (-0.0001 to 0.0019) | 0.0845 | **0.1394 (0.0676 to 0.2112)** | **0.0001** | 5073 |
| Numeric memory Instance 2 | 0.0007 (-0.0002 to 0.0015) | 0.1292 | 0.0608 (-0.0072 to 0.1289) | 0.0797 | 6617 |
| Numeric memory Instance 3 | 0.0004 (-0.0020 to 0.0028) | 0.7322 | 0.0933 (-0.0983 to 0.2850) | 0.3393 | 751 |
| Prospective memory result Instance 0 | 0.0001 (-0.0001 to 0.0002) | 0.5375 | 0.0070 (-0.0048 to 0.0189) | 0.2451 | 17846 |
| Prospective memory result Instance 2 | 0.00001 (-0.0002 to 0.0002) | 0.9107 | 0.0072 (-0.0085 to 0.0230) | 0.3699 | 11578 |
| **Prospective memory result Instance 3** | 0.0005 (-0.0001 to 0.0011) | 0.1206 | **0.0586 (0.0096 to 0.1076)** | **0.0190** | 1021 |

*Table S. 4 Associations between MVPA and cognitive function using fully adjusted model. Bold results show significant (p<0.05) results.*

| Fully Adjusted Model | MVPA in mins per week | | WHO Guidelines | | |
|---|---|---|---|---|---|
| Cognitive Variables | **Estimate (95% CI)** | **P-value** | **Estimate (95% CI)** | **P-value** | N_Obs |
| Reaction Time Instance 0 | 0.0081 (-0.0228 to 0.0390) | 0.6078 | -0.0639 (-2.2435 to 2.1158) | 0.9542 | 45784 |
| Reaction Time Instance 2 | 0.0416 (-0.0196 to 0.1029) | 0.1825 | -2.7844 (-7.4653 to 1.8965) | 0.2436 | 11520 |
| Reaction Time Instance 3 | 0.1549 (-0.0491 to 0.3590) | 0.1366 | -7.1781 (-23.0184 to 8.6623) | 0.3741 | 1002 |
| Fluid Intelligence Instance 0 | 0.0003 (-0.0006 to 0.0012) | 0.5027 | 0.0340 (-0.0276 to 0.0957) | 0.2793 | 17651 |
| Fluid Intelligence Instance 2 | 0.0004 (-0.0006 to 0.0015) | 0.4018 | -0.0042 (-0.0841 to 0.0756) | 0.9176 | 11351 |
| Fluid Intelligence Instance 3 | 0.0030 (-0.0004 to 0.0065) | 0.0855 | 0.1874 (-0.0794 to 0.4542) | 0.1684 | 1001 |
| Trail making test 1Instance 2 | -0.0206 (-0.0781 to 0.0370) | 0.4839 | 3.2902 (-1.1782 to 7.7585) | 0.1489 | 8546 |
| Trail making test 1Instance 3 | 0.1285 (-0.0687 to 0.3257) | 0.2012 | -0.5031 (-15.7874 to 14.7811) | 0.9485 | 998 |
| Trail making test 2 Instance 2 | 0.0477 (-0.1255 to 0.2210) | 0.5890 | 0.0043 (-13.5146 to 13.5233) | 0.9995 | 8302 |
| Trail making test 2 Instance 3 | 0.0858 (-0.4335 to 0.6051) | 0.7459 | -4.4148 (-44.6735 to 35.8439) | 0.8297 | 976 |
| Paired associate learning Instance 2 | -0.0004 (-0.0020 to 0.0012) | 0.6449 | 0.0990 (-0.0241 to 0.2220) | 0.1150 | 8646 |
| **Paired associate learning Instance 3** | 0.0031 (-0.0015 to 0.0077) | 0.1906 | **0.4036 (0.0444 to 0.7627)** | **0.0277** | **1021** |
| Numeric memory Instance 0 | -0.0003 (-0.0013 to 0.0007) | 0.5446 | 0.0498 (-0.0245 to 0.1241) | 0.1887 | 5073 |
| Numeric memory Instance 2 | -0.0006 (-0.0015 to 0.0003) | 0.2035 | -0.0405 (-0.1096 to 0.0287) | 0.2511 | 6617 |
| Numeric memory Instance 3 | -0.0002 (-0.0028 to 0.0024) | 0.8871 | 0.0725 (-0.1271 to 0.2721) | 0.4759 | 751 |
| Prospective memory result 0 | -0.00008 (-0.0003 to 0.00008) | 0.3244 | -0.0024 (-0.0146 to 0.0098) | 0.6985 | 17846 |
| Prospective memory result 2 | -0.00017 (-0.0004 to 0.00004) | 0.1071 | -0.0059 (-0.0221 to 0.0102) | 0.4728 | 11578 |
| Prospective memory result 3 | 0.000271 (-0.0004 to 0.0009) | 0.4138 | 0.0463 (-0.0044 to 0.0970) | 0.0735 | 1021 |

*Table S. 5 Associations between MVPA and average brain region volumes using based model (MVPA in minutes per week). Bold results show significant (p<0.05) results.*

| Age and Sex Adjusted Model | MVPA in mins per week. | | WHO based guidelines | | N_Obs |
|---|---|---|---|---|---|
| Cognitive_Var | Estimate (95% CI) | P-value | Estimate (95% CI) | P-value | |

| | | | | | |
|---|---|---|---|---|---|
| AvgVolumeHippocampusInst2 | 0.2481 (-0.0085 to 0.5047) | 0.0581 | **20.6136 (1.1897 to 40.0375)** | **0.0375** | **8387** |
| AvgVolumeHippocampusInst3 | -0.2240 (-1.0666 to 0.6187) | 0.6019 | -15.5487 (-81.2263 to 50.1290) | 0.6422 | 745 |
| AvgVolumeAccumbensInst2 | **0.1311 (0.0722 to 0.1899)** | **<0.0001** | **8.0802 (3.6251 to 12.5353)** | **0.0004** | **8387** |
| AvgVolumeAccumbensInst3 | 0.0956 (-0.0888 to 0.2799) | 0.3091 | 4.5703 (-9.8022 to 18.9429) | 0.5326 | 745 |
| AvgVolumeAmygdalaInst2 | 0.1182 (-0.0114 to 0.2478) | 0.0738 | **11.9073 (2.1006 to 21.7140)** | **0.0173** | **8387** |
| AvgVolumeAmygdalaInst3 | 0.3544 (-0.0543 to 0.7630) | 0.0891 | **36.5238 (4.7221 to 68.3255)** | **0.0244** | **745** |
| AvgVolumeCaudateInst2 | **0.5431 (0.2936 to 0.7927)** | **<0.0001** | **39.9644 (21.0754 to 58.8534)** | **0.00003** | **8387** |
| AvgVolumeCaudateInst3 | 0.5701 (-0.2056 to 1.3458) | 0.1495 | -18.4615 (-78.9915 to 42.0685) | 0.5495 | 745 |
| AvgVolumePallidumInst2 | **0.4232 (0.2798 to 0.5666)** | **<0.0001** | **30.5261 (19.6688 to 41.3835)** | **<0.00001** | **8387** |
| AvgVolumePallidumInst3 | 0.3123 (-0.1459 to 0.7704) | 0.1812 | 13.9512 (-21.7856 to 49.6881) | 0.4437 | 745 |
| AvgVolumePutamenInst2 | **0.7297 (0.4237 to 1.0357)** | **<0.0001** | **50.2300 (27.0623 to 73.3977)** | **0.00002** | **8387** |
| AvgVolumePutamenInst3 | **1.1717 (0.1966 to 2.1467)** | **0.0186** | 21.9688 (-54.2946 to 98.2322) | 0.5719 | 745 |
| AvgVolumeThalamusInst2 | **0.9380 (0.5554 to 1.3206)** | **<0.0001** | **58.5992 (29.6240 to 87.5743)** | **0.00007** | **8387** |
| AvgVolumeThalamusInst3 | 0.7339 (-0.4315 to 1.8993) | 0.2168 | -18.5648 (-109.4800 to 72.3499) | 0.6886 | 745 |

*Table S. 6 Associations between MVPA and average brain region volumes using fully adjusted model (MVPA in minutes per week). Bold results show significant (p<0.05) results.*

| Cognitive_Var | MVPA in mins per week. | | WHO based guidelines | | |
|---|---|---|---|---|---|
| | Estimate (95% CI) | P-value | Estimate (95% CI) | P-value | N_Obs |
| AvgVolumeHippocampusInst2 | 0.06006 (-0.2071 to 0.3272) | 0.6594 | 6.697 (-13.417 to 26.8120) | 0.5140 | 8387 |
| AvgVolumeHippocampusInst3 | -0.2529 (-1.1449 to 0.6390) | 0.5779 | -15.093 (-83.951 to 53.7644) | 0.6671 | 745 |
| AvgVolumeAccumbensInst2 | 0.08612 (0.0251 to 0.1471) | 0.0056 | 4.5351 (-0.058 to 9.1280) | 0.0530 | 8387 |
| AvgVolumeAccumbensInst3 | 0.09514 (-0.1004 to 0.2907) | 0.3398 | 4.3901 (-10.712 to 19.4918) | 0.5684 | 745 |
| AvgVolumeAmygdalaInst2 | 0.08618 (-0.0490 to 0.2214) | 0.2117 | 10.0722 (-0.111 to 20.2550) | 0.0525 | 8387 |
| AvgVolumeAmygdalaInst3 | 0.2988 (-0.1340 to 0.7317) | 0.1757 | 35.457 (2.102 to 68.8137) | 0.0372 | 745 |
| AvgVolumeCaudateInst2 | 0.2667 (0.0077 to 0.5257) | 0.0435 | 19.878 (0.381 to 39.3758) | 0.0457 | 8387 |
| AvgVolumeCaudateInst3 | 0.3950 (-0.4253 to 1.2153) | 0.3448 | -33.219 (-96.533 to 30.0929) | 0.3033 | 745 |
| AvgVolumePallidumInst2 | 0.3031 (0.1539 to 0.4524) | 0.0001 | 21.004 (9.765 to 32.2429) | 0.0003 | 8387 |
| AvgVolumePallidumInst3 | 0.2261 (-0.2615 to 0.7138) | 0.3629 | 6.549 (-31.113 to 44.211) | 0.7329 | 745 |
| AvgVolumePutamenInst2 | 0.3720 (0.0546 to 0.6894) | 0.0216 | 23.254 (-0.646 to 47.1549) | 0.0565 | 8387 |
| AvgVolumePutamenInst3 | 0.8825 (-0.1453 to 1.9104) | 0.0923 | 4.326 (-75.171 to 83.8240) | 0.9149 | 745 |
| AvgVolumeThalamusInst2 | 0.5418 (0.1461 to 0.9376) | 0.0072 | 27.814 (-1.989 to 57.6182) | 0.0674 | 8387 |
| AvgVolumeThalamusInst3 | 0.3837 (-0.8484 to 1.6159) | 0.5410 | -47.479 (-142.551 to 47.5924) | 0.3272 | 745 |

*Table S. 7 Associations between MVPA and average brain region volumes using Base model (MVPA in minutes per week). Bold results show significant (p<0.05) results.*

| Cognitive_Var | Estimate | P_Value | Lower_CI | Upper_CI | N_Obs |
|---|---|---|---|---|---|
| VolumeHippocampusLeftInst2 | 0.1446 | 0.3171 | -0.1387 | 0.4280 | 8387 |
| VolumeHippocampusLeftInst3 | -0.2816 | 0.5392 | -1.1815 | 0.6183 | 745 |
| VolumeHippocampusRightInst2 | 0.3515 | 0.0177 | 0.0609 | 0.6422 | 8387 |
| VolumeHippocampusRightInst3 | -0.1664 | 0.7370 | -1.1386 | 0.8058 | 745 |
| **VolumeAccumbensLeftInst2** | **0.1233** | **0.0005** | **0.0539** | **0.1926** | **8387** |
| VolumeAccumbensLeftInst3 | 0.1023 | 0.3590 | -0.1165 | 0.3212 | 745 |
| **VolumeAccumbensRightInst2** | **0.1389** | **<0.0001** | **0.0748** | **0.2030** | **8387** |
| VolumeAccumbensRightInst3 | 0.0888 | 0.3928 | -0.1151 | 0.2927 | 745 |

| | | | | | |
|---|---|---|---|---|---|
| **VolumeAmygdalaLeftInst2** | **0.1653** | **0.0323** | **0.0139** | **0.3167** | **8387** |
| VolumeAmygdalaLeftInst3 | 0.4662 | 0.0535 | -0.0071 | 0.9395 | 745 |
| VolumeAmygdalaRightInst2 | 0.0711 | 0.3994 | -0.0943 | 0.2366 | 8387 |
| VolumeAmygdalaRightInst3 | 0.2425 | 0.3615 | -0.2789 | 0.7640 | 745 |
| **VolumeCaudateLeftInst2** | **0.5336** | **<0.0001** | **0.2835** | **0.7836** | **8387** |
| VolumeCaudateLeftInst3 | 0.5135 | 0.1951 | -0.2638 | 1.2908 | 745 |
| **VolumeCaudateRightInst2** | **0.5527** | **<0.0001** | **0.2860** | **0.8195** | **8387** |
| VolumeCaudateRightInst3 | 0.6267 | 0.1389 | -0.2038 | 1.4572 | 745 |
| **VolumePallidumLeftInst2** | **0.3546** | **<0.0001** | **0.2000** | **0.5092** | **8387** |
| VolumePallidumLeftInst3 | 0.2582 | 0.3099 | -0.2406 | 0.7569 | 745 |
| **VolumePallidumRightInst2** | **0.4918** | **<0.0001** | **0.3407** | **0.6430** | **8387** |
| VolumePallidumRightInst3 | 0.3664 | 0.1373 | -0.1171 | 0.8500 | 745 |
| **VolumePutamenLeftInst2** | **0.6886** | **<0.0001** | **0.3576** | **1.0197** | **8387** |
| **VolumePutamenLeftInst3** | **1.2220** | **0.0214** | **0.1813** | **2.2627** | **745** |
| **VolumePutamenRightInst2** | **0.7708** | **<0.0001** | **0.4489** | **1.0926** | **8387** |
| **VolumePutamenRightInst3** | **1.1214** | **0.0339** | **0.0856** | **2.1571** | **745** |
| **VolumeThalamusLeftInst2** | **0.9056** | **<0.0001** | **0.5062** | **1.3050** | **8387** |
| VolumeThalamusLeftInst3 | 0.8371 | 0.1764 | -0.3772 | 2.0514 | 745 |
| **VolumeThalamusRightInst2** | **0.9704** | **<0.0001** | **0.5880** | **1.3528** | **8387** |
| VolumeThalamusRightInst3 | 0.6307 | 0.2924 | -0.5445 | 1.8059 | 745 |

*Table S. 8 Associations between MVPA and average brain region volumes using Base model (categorical MVPA based on WHO guidelines). Bold results show significant (p<0.05) results.*

| Cognitive_Var | Estimate | P_Value | Lower_CI | Upper_CI | N_Obs |
|---|---|---|---|---|---|
| VolumeHippocampusLeftInst2 | 16.569 | 0.130 | -4.879 | 38.016 | 8387 |
| VolumeHippocampusLeftInst3 | -27.835 | 0.436 | -97.961 | 42.291 | 745 |
| **VolumeHippocampusRightInst2** | **24.659** | **0.028** | **2.660** | **46.657** | **8387** |
| VolumeHippocampusRightInst3 | -3.262 | 0.933 | -79.038 | 72.514 | 745 |
| **VolumeAccumbensLeftInst2** | **7.290** | **0.007** | **2.040** | **12.541** | **8387** |
| VolumeAccumbensLeftInst3 | 4.145 | 0.634 | -12.919 | 21.208 | 745 |
| **VolumeAccumbensRightInst2** | **8.870** | **<0.0001** | **4.015** | **13.725** | **8387** |
| VolumeAccumbensRightInst3 | 4.996 | 0.537 | -10.898 | 20.890 | 745 |
| **VolumeAmygdalaLeftInst2** | **14.622** | **0.012** | **3.164** | **26.079** | **8387** |
| VolumeAmygdalaLeftInst3 | 25.164 | 0.181 | -11.772 | 62.101 | 745 |
| VolumeAmygdalaRightInst2 | 9.193 | 0.150 | -3.330 | 21.716 | 8387 |
| **VolumeAmygdalaRightInst3** | **47.883** | **0.021** | **7.367** | **88.399** | **745** |
| **VolumeCaudateLeftInst2** | **40.209** | **<0.0001** | **21.282** | **59.136** | **8387** |
| VolumeCaudateLeftInst3 | -13.641 | 0.659 | -74.285 | 47.003 | 745 |
| **VolumeCaudateRightInst2** | **39.720** | **<0.0001** | **19.527** | **59.913** | **8387** |

| | | | | | |
|---|---|---|---|---|---|
| VolumeCaudateRightInst3 | -23.282 | 0.481 | -88.082 | 41.519 | 745 |
| **VolumePallidumLeftInst2** | **29.807** | **<0.0001** | **18.107** | **41.507** | **8387** |
| VolumePallidumLeftInst3 | 23.378 | 0.238 | -15.484 | 62.241 | 745 |
| **VolumePallidumRightInst2** | **31.245** | **<0.0001** | **19.797** | **42.694** | **8387** |
| VolumePallidumRightInst3 | 4.524 | 0.814 | -33.220 | 42.268 | 745 |
| **VolumePutamenLeftInst2** | **50.288** | **<0.0001** | **25.224** | **75.352** | **8387** |
| VolumePutamenLeftInst3 | 11.693 | 0.778 | -69.705 | 93.092 | 745 |
| **VolumePutamenRightInst2** | **50.172** | **<0.0001** | **25.801** | **74.544** | **8387** |
| VolumePutamenRightInst3 | 32.245 | 0.434 | -48.692 | 113.181 | 745 |
| **VolumeThalamusLeftInst2** | **51.381** | **0.001** | **21.135** | **81.627** | **8387** |
| VolumeThalamusLeftInst3 | -15.986 | 0.741 | -110.736 | 78.763 | 745 |
| **VolumeThalamusRightInst2** | **65.818** | **<0.0001** | **36.866** | **94.769** | **8387** |
| VolumeThalamusRightInst3 | -21.143 | 0.651 | -112.793 | 70.507 | 745 |

*Table S. 9. Associations between MVPA and average brain region volumes using fully adjusted model (MVPA in minutes per week). Bold results show significant (p<0.05) results.*

| Cognitive_Var | Estimate | P_Value | Lower_CI | Upper_CI | N_Obs |
|---|---|---|---|---|---|
| VolumeHippocampusLeftInst2 | -0.0753 | 0.6168 | -0.3704 | 0.2198 | 8387 |
| VolumeHippocampusLeftInst3 | -0.3712 | 0.4448 | -1.3246 | 0.5821 | 745 |
| VolumeHippocampusRightInst2 | 0.1955 | 0.2059 | -0.1074 | 0.4984 | 8387 |
| VolumeHippocampusRightInst3 | -0.1346 | 0.7974 | -1.1634 | 0.8942 | 745 |
| **VolumeAccumbensLeftInst2** | **0.0767** | **0.0370** | **0.0046** | **0.1487** | **8387** |
| VolumeAccumbensLeftInst3 | 0.0760 | 0.5213 | -0.1565 | 0.3085 | 745 |
| **VolumeAccumbensRightInst2** | **0.0955** | **0.0049** | **0.0290** | **0.1621** | **8387** |
| VolumeAccumbensRightInst3 | 0.1143 | 0.3001 | -0.1021 | 0.3307 | 745 |
| VolumeAmygdalaLeftInst2 | 0.1466 | 0.0690 | -0.0114 | 0.3046 | 8387 |
| VolumeAmygdalaLeftInst3 | 0.3436 | 0.1796 | -0.1586 | 0.8458 | 745 |
| VolumeAmygdalaRightInst2 | 0.0258 | 0.7702 | -0.1471 | 0.1987 | 8387 |
| VolumeAmygdalaRightInst3 | 0.2541 | 0.3668 | -0.2982 | 0.8063 | 745 |
| **VolumeCaudateLeftInst2** | **0.2771** | **0.0365** | **0.0173** | **0.5368** | **8387** |
| VolumeCaudateLeftInst3 | 0.3752 | 0.3704 | -0.4467 | 1.1971 | 745 |
| VolumeCaudateRightInst2 | 0.2564 | 0.0695 | -0.0204 | 0.5333 | 8387 |
| VolumeCaudateRightInst3 | 0.4148 | 0.3544 | -0.4639 | 1.2935 | 745 |
| **VolumePallidumLeftInst2** | **0.2445** | **0.0029** | **0.0834** | **0.4056** | **8387** |
| VolumePallidumLeftInst3 | 0.1663 | 0.5382 | -0.3639 | 0.6966 | 745 |
| **VolumePallidumRightInst2** | **0.3619** | **<0.0001** | **0.2046** | **0.5192** | **8387** |
| VolumePallidumRightInst3 | 0.2860 | 0.2751 | -0.2280 | 0.8000 | 745 |
| VolumePutamenLeftInst2 | 0.3433 | 0.0503 | -0.0005 | 0.6872 | 8387 |
| VolumePutamenLeftInst3 | 1.0466 | 0.0622 | -0.0536 | 2.1468 | 745 |
| **VolumePutamenRightInst2** | **0.4007** | **0.0188** | **0.0666** | **0.7349** | **8387** |
| VolumePutamenRightInst3 | 0.7185 | 0.1957 | -0.3706 | 1.8076 | 745 |

| Cognitive_Var | Estimate | P_Value | Lower_CI | Upper_CI | N_Obs |
| --- | --- | --- | --- | --- | --- |
| **VolumeThalamusLeftInst2** | **0.4850** | **0.0214** | **0.0719** | **0.8981** | **8387** |
| VolumeThalamusLeftInst3 | 0.4573 | 0.4853 | -0.8289 | 1.7436 | 745 |
| **VolumeThalamusRightInst2** | **0.5988** | **0.0030** | **0.2029** | **0.9946** | **8387** |
| VolumeThalamusRightInst3 | 0.3102 | 0.6237 | -0.9305 | 1.5509 | 745 |

*Table S. 10. Associations between MVPA and average brain region volumes using Fully adjusted model (categorical MVPA based on WHO guidelines). Bold results show significant (p<0.05) results.*

| Cognitive_Var | Estimate | P_Value | Lower_CI | Upper_CI | N_Obs |
| --- | --- | --- | --- | --- | --- |
| VolumeHippocampusLeftInst2 | 0.9850 | 0.9308 | -21.2339 | 23.2039 | 8387 |
| VolumeHippocampusLeftInst3 | -33.4081 | 0.3730 | -106.9861 | 40.1700 | 745 |
| VolumeHippocampusRightInst2 | 12.4100 | 0.2862 | -10.3967 | 35.2167 | 8387 |
| VolumeHippocampusRightInst3 | 3.2217 | 0.9365 | -76.1924 | 82.6358 | 745 |
| VolumeAccumbensLeftInst2 | 3.7022 | 0.1811 | -1.7233 | 9.1277 | 8387 |
| VolumeAccumbensLeftInst3 | 2.3951 | 0.7934 | -15.5532 | 20.3433 | 745 |
| **VolumeAccumbensRightInst2** | **5.3681** | **0.0359** | **0.3544** | **10.3817** | **8387** |
| VolumeAccumbensRightInst3 | 6.3851 | 0.4533 | -10.3215 | 23.0918 | 745 |
| **VolumeAmygdalaLeftInst2** | **13.4481** | **0.0267** | **1.5550** | **25.3412** | **8387** |
| VolumeAmygdalaLeftInst3 | 19.1692 | 0.3322 | -19.6169 | 57.9552 | 745 |
| VolumeAmygdalaRightInst2 | 6.6962 | 0.3133 | -6.3202 | 19.7126 | 8387 |
| **VolumeAmygdalaRightInst3** | **51.7464** | **0.0170** | **9.2603** | **94.2325** | **745** |
| **VolumeCaudateLeftInst2** | **21.6840** | **0.0298** | **2.1288** | **41.2393** | **8387** |
| VolumeCaudateLeftInst3 | -24.3431 | 0.4516 | -87.7970 | 39.1109 | 745 |
| VolumeCaudateRightInst2 | 18.0730 | 0.0892 | -2.7719 | 38.9180 | 8387 |
| VolumeCaudateRightInst3 | -42.0966 | 0.2232 | -109.8954 | 25.7022 | 745 |
| **VolumePallidumLeftInst2** | **21.3795** | **0.0006** | **9.2533** | **33.5057** | **8387** |
| VolumePallidumLeftInst3 | 15.7331 | 0.4506 | -25.1912 | 56.6574 | 745 |
| **VolumePallidumRightInst2** | **20.6286** | **0.0006** | **8.7797** | **32.4775** | **8387** |
| VolumePallidumRightInst3 | -2.6340 | 0.8964 | -42.3404 | 37.0723 | 745 |
| VolumePutamenLeftInst2 | 24.7401 | 0.0611 | -1.1477 | 50.6279 | 8387 |
| VolumePutamenLeftInst3 | 0.2379 | 0.9956 | -84.8909 | 85.3667 | 745 |
| VolumePutamenRightInst2 | 21.7686 | 0.0899 | -3.3937 | 46.9309 | 8387 |
| VolumePutamenRightInst3 | 8.4153 | 0.8444 | -75.7455 | 92.5760 | 745 |
| VolumeThalamusLeftInst2 | 18.4555 | 0.2449 | -12.6542 | 49.5653 | 8387 |
| VolumeThalamusLeftInst3 | -46.6946 | 0.3560 | -145.9493 | 52.5602 | 745 |
| **VolumeThalamusRightInst2** | **37.1739** | **0.0145** | **7.3659** | **66.9819** | **8387** |
| VolumeThalamusRightInst3 | -48.2641 | 0.3226 | -143.9846 | 47.4564 | 745 |

*Table S. 11 Sensitivity analysis for different MVPA mins categories (less than 50 is reference category) – Subcortical regions*

| Brain regions | 51-100 | | 101-150 | | MVPA_group 151-300 | | MVPA_group>300 | | N_Obs |
|---|---|---|---|---|---|---|---|---|---|
| | Estimate_CI | P_Value | Estimate_CI | P_Value | Estimate_CI | P_Value | Estimate_CI | P_Value | N_Obs |
| VolumeHippocampusInst2 | 20.3810 (-19.8004 to 60.5623) | 0.3201 | 33.6491 (-5.9631 to 73.2613) | 0.0959 | 23.7015 (-10.3661 to 57.7692) | 0.1727 | 28.6792 (-5.1692 to 62.5275) | 0.0968 | 8387 |
| VolumeHippocampusInst3 | 94.1628 (-50.3388 to 238.6644) | 0.2012 | 125.5606 (-11.8367 to 262.9580) | 0.0732 | 95.7923 (-27.2054 to 218.7899) | 0.1267 | 46.9463 (-73.4035 to 167.2962) | 0.4440 | 745 |
| **VolumeAccumbensInst2** | 5.9626 (-3.2112 to 15.1364) | 0.2027 | **9.3820 (0.3381 to 18.4258)** | **0.0420** | **8.8497 (1.0718 to 16.6277)** | **0.0257** | **11.1985 (3.4706 to 18.9264)** | **0.0045** | **8387** |
| **VolumeAccumbensInst3** | 35.1595 (3.5571 to 66.7619) | 0.0293 | **44.2846 (14.2359 to 74.3334)** | **0.0039** | **31.3937 (4.4942 to 58.2933)** | **0.0222** | **36.6721 (10.3517 to 62.9926)** | **0.0064** | **745** |
| VolumeAmygdalaInst2 | 5.0635 (-15.2785 to 25.4054) | 0.6256 | -10.0098 (-30.0636 to 10.0440) | 0.3279 | 6.6368 (-10.6100 to 23.8837) | 0.4507 | 9.0514 (-8.0844 to 26.1872) | 0.3005 | 8387 |
| VolumeAmygdalaInst3 | -11.2317 (-81.4437 to 58.9803) | 0.7536 | 1.6514 (-65.1086 to 68.4115) | 0.9613 | 25.2712 (-34.4922 to 85.0346) | 0.4067 | 37.8243 (-20.6525 to 96.3012) | 0.2045 | 745 |
| **VolumeCaudateInst2** | 10.6774 (-28.2750 to 49.6298) | 0.5911 | 17.9026 (-20.4980 to 56.3032) | 0.3608 | 26.6407 (-6.3850 to 59.6664) | 0.1139 | **33.4055 (0.5924 to 66.2185)** | **0.0460** | 8387 |
| VolumeCaudateInst3 | -33.7954 (-166.7088 to 99.1180) | 0.6178 | 67.5253 (-58.8535 to 193.9042) | 0.2945 | -41.5602 (-154.6941 to 71.5737) | 0.4710 | 2.3320 (-108.3665 to 113.0304) | 0.9670 | 745 |
| **VolumePallidumInst2** | -11.9236 (-34.3672 to 10.5199) | 0.2977 | 15.7243 (-6.4014 to 37.8499) | 0.1636 | 17.9873 (-1.0414 to 37.0160) | 0.0639 | **27.2734 (8.3672 to 46.1796)** | **0.0047** | 8387 |
| VolumePallidumInst3 | 50.1522 (-29.0722 to 129.3767) | 0.2143 | 31.4868 (-43.8427 to 106.8163) | 0.4121 | 34.0202 (-33.4145 to 101.4549) | 0.3223 | 38.0064 (-27.9766 to 103.9895) | 0.2585 | 745 |
| **VolumePutamenInst2** | 17.6798 (-30.0501 to 65.4096) | 0.4678 | **61.0994 (14.0457 to 108.1532)** | **0.0109** | **44.8900 (4.4224 to 85.3576)** | **0.0297** | **59.0198 (18.8127 to 99.2269)** | **0.0040** | 8387 |
| **VolumePutamenInst3** | 125.8285 (-40.8089 to 292.4659) | 0.1387 | **208.4797 (50.0348 to 366.9246)** | **0.0100** | 136.7847 (-5.0546 to 278.6240) | 0.0587 | 128.8303 (-9.9557 to 267.6162) | 0.0688 | 745 |
| **VolumeThalamusInst2** | 22.6560 (-36.8716 to 82.1835) | 0.4557 | **67.6493 (8.9650 to 126.3337)** | **0.0239** | **58.6813 (8.2110 to 109.1515)** | **0.0227** | **63.4698 (13.3244 to 113.6151)** | **0.0131** | 8387 |
| VolumeThalamusInst3 | 160.4179 (-39.2361 to 360.0718) | 0.1151 | 183.5588 (-6.2794 to 373.3970) | 0.0581 | 73.6555 (-96.2870 to 243.5980) | 0.3951 | 87.9469 (-78.3372 to 254.2311) | 0.2995 | 745 |

*Table S. 12 Associations between MVPA and cognitive function using fully model (MVPA in minutes per week) stratified by age. Bold results show significant (p<0.05) results.*

| MVPA | Below 65 years | | | Above 65 years | | |
|---|---|---|---|---|---|---|
| Cognitive_Var | Estimate_CI | P_Value | N_Obs | Estimate_CI | P_Value | N_Obs |
| ReactionTime0 | -0.0071 (-0.0349 to 0.0208) | 0.6187 | 40622 | -0.0259 (-0.0647 to 0.0129) | 0.1910 | 31679 |
| ReactionTime2 | 0.0202 (-0.0322 to 0.0725) | 0.4500 | 12797 | 0.0092 (-0.0718 to 0.0903) | 0.8231 | 7160 |
| ReactionTime3 | 0.0900 (-0.0576 to 0.2375) | 0.2322 | 1505 | -0.0092 (-0.3045 to 0.2860) | 0.9512 | 561 |
| FluidIntelligence0 | 0.0008 (-0.0001 to 0.0017) | 0.0832 | 16203 | 0.0004 (-0.0006 to 0.0015) | 0.4233 | 12254 |
| FluidIntelligence2 | 0.0003 (-0.0007 to 0.0013) | 0.5645 | 12742 | 0.0005 (-0.0008 to 0.0019) | 0.4220 | 7023 |
| FluidIntelligence3 | 0.0012 (-0.0016 to 0.0041) | 0.4009 | 1502 | 0.0040 (-0.0007 to 0.0088) | 0.0979 | 559 |

| | | | | | | |
|---|---|---|---|---|---|---|
| DurationNumericPtrail1Ins2 | 0.0170 (-0.0225 to 0.0564) | 0.3998 | 9990 | -0.0634 (-0.1446 to 0.0178) | 0.1262 | 5201 |
| DurationNumericPtrail1Ins3 | 0.0530 (-0.0443 to 0.1503) | 0.2861 | 1505 | 0.1254 (-0.2105 to 0.4614) | 0.4647 | 557 |
| DurationAlphaNumePtrail2Ins2 | 0.1048 (-0.0128 to 0.2225) | 0.0807 | 9895 | -0.0851 (-0.3340 to 0.1638) | 0.5028 | 5004 |
| DurationAlphaNumePtrail2Ins3 | -0.0018 (-0.2829 to 0.2792) | 0.9897 | 1495 | 0.2344 (-0.5825 to 1.0514) | 0.5741 | 539 |
| NumberWordPairsCorrectAssociated2 | -0.0008 (-0.0022 to 0.0006) | 0.2683 | 10044 | 0.0003 (-0.0018 to 0.0024) | 0.7559 | 5281 |
| NumberWordPairsCorrectAssociated3 | 0.0021 (-0.0014 to 0.0055) | 0.2454 | 1523 | 0.0012 (-0.0054 to 0.0077) | 0.7282 | 575 |
| MaxDigitRememCorrectly0 | -0.0003 (-0.0014 to 0.0007) | 0.5272 | 4523 | 0.0001 (-0.0011 to 0.0013) | 0.8551 | 3536 |
| MaxDigitRememCorrectly2 | 0.0001 (-0.0008 to 0.0009) | 0.9000 | 7358 | -0.0009 (-0.0020 to 0.0003) | 0.1417 | 4120 |
| MaxDigitRememCorrectly3 | -0.0006 (-0.0027 to 0.0015) | 0.5930 | 1174 | 0.0015 (-0.0021 to 0.0050) | 0.4185 | 420 |
| ProspectiveMemR0 | -0.0008 (-0.0022 to 0.0006) | 0.2630 | 16315 | -0.0004 (-0.0019 to 0.0011) | 0.6039 | 12399 |
| ProspectiveMemR1 | -0.0003 (-0.0039 to 0.0033) | 0.8517 | 3197 | 0.0003 (-0.0027 to 0.0033) | 0.8662 | 3236 |
| ProspectiveMemR3 | 0.0031 (-0.0022 to 0.0087) | 0.2663 | 1523 | 0.0045 (-0.0026 to 0.0119) | 0.2274 | 575 |

*Table S. 13 Associations between MVPA and cognitive function using fully model (categorical MVPA based on WHO guidelines) stratified by age. Bold results show significant (p<0.05) results.*

| | Below 65 years | | | Above 65 years | | |
|---|---|---|---|---|---|---|
| Cognitive_Var | Estimate_CI | P_Value | N_Obs | Estimate_CI | P_Value | N_Obs |
| ReactionTime0 | -0.9389 (-3.0672 to 1.1893) | 0.3872 | 40625 | -2.0460 (-4.6913 to 0.5992) | 0.1295 | 31680 |
| ReactionTime2 | -0.8715 (-4.9410 to 3.1979) | 0.6747 | 12798 | -5.0687 (-11.0947 to 0.9572) | 0.0993 | 7160 |
| ReactionTime3 | -3.5692 (-15.0909 to 7.9526) | 0.5438 | 1505 | -15.7264 (-37.1034 to 5.6507) | 0.1499 | 561 |
| **FluidIntelligence0** | **0.1315 (0.0630 to 0.2001)** | **0.0002** | **16204** | 0.0191 (-0.0533 to 0.0915) | 0.6057 | 12254 |
| FluidIntelligence2 | -0.0276 (-0.1049 to 0.0497) | 0.4846 | 12743 | 0.0185 (-0.0812 to 0.1182) | 0.7166 | 7023 |
| FluidIntelligence3 | 0.0526 (-0.1695 to 0.2748) | 0.6424 | 1502 | 0.3433 (-0.0017 to 0.6882) | 0.0517 | 559 |
| **DurationNumericPtrail1Ins2** | **4.3057 (1.1921 to 7.4193)** | **0.0067** | **9991** | 0.3336 (-5.8328 to 6.5000) | 0.9156 | 5201 |
| DurationNumericPtrail1Ins3 | 6.6093 (-0.9906 to 14.2092) | 0.0885 | 1505 | -7.7199 (-31.9731 to 16.5333) | 0.5330 | 557 |
| DurationAlphaNumePtrail2Ins2 | 8.9423 (-0.3604 to 18.2450) | 0.0596 | 9896 | -14.3124 (-33.2915 to 4.6666) | 0.1395 | 5004 |
| DurationAlphaNumePtrail2Ins3 | 10.3281 (-11.6469 to 32.3030) | 0.3571 | 1495 | -33.0460 (-91.7711 to 25.6790) | 0.2706 | 539 |
| NumberWordPairsCorrectAssociated2 | 0.0439 (-0.0657 to 0.1535) | 0.4322 | 10045 | 0.1093 (-0.0511 to 0.2698) | 0.1818 | 5281 |
| NumberWordPairsCorrectAssociated3 | 0.1613 (-0.1089 to 0.4315) | 0.2422 | 1523 | 0.3409 (-0.1405 to 0.8222) | 0.1657 | 575 |
| MaxDigitRememCorrectly0 | 0.0123 (-0.0696 to 0.0941) | 0.7690 | 4523 | 0.0531 (-0.0347 to 0.1409) | 0.2363 | 3536 |
| MaxDigitRememCorrectly2 | 0.0115 (-0.0532 to 0.0763) | 0.7271 | 7359 | -0.0626 (-0.1505 to 0.0253) | 0.1629 | 4120 |
| MaxDigitRememCorrectly3 | -0.0102 (-0.1767 to 0.1564) | 0.9047 | 1174 | 0.1036 (-0.1494 to 0.3567) | 0.4227 | 420 |
| ProspectiveMemR0 | -0.0310 (-0.1372 to 0.0743) | 0.5655 | 16316 | -0.0080 (-0.1086 to 0.0921) | 0.8757 | 12399 |
| ProspectiveMemR1 | 0.0131 (-0.2566 to 0.2775) | 0.9234 | 3197 | 0.0524 (-0.1632 to 0.2658) | 0.6320 | 3236 |
| ProspectiveMemR3 | 0.1088 (-0.2873 to 0.4935) | 0.5843 | 1523 | **0.6651 (0.1937 to 1.1345)** | **0.0055** | **575** |

*Table S. 14 Associations between MVPA and cognitive function using fully model (MVPA in minutes per week) stratified by sex. Bold results show significant (p<0.05) results.*

|  | Males | | | Females | | |
|---|---|---|---|---|---|---|
| Cognitive_Var | Estimate_CI | P_Value | N_Obs | Estimate_CI | P_Value | N_Obs |
| ReactionTime0 | 0.0099 (-0.0289 to 0.0487) | 0.6163 | 21463 | 0.0039 (-0.0468 to 0.0545) | 0.8812 | 24321 |
| **ReactionTime2** | **0.0854 (0.0079 to 0.1629)** | **0.0309** | **5930** | -0.0416 (-0.1419 to 0.0588) | 0.4170 | 5590 |
| ReactionTime3 | 0.2191 (-0.0435 to 0.4817) | 0.1027 | 525 | 0.0737 (-0.2506 to 0.3981) | 0.6562 | 477 |
| FluidIntelligence0 | 0.0004 (-0.0008 to 0.0016) | 0.4801 | 8287 | 0.0000 (-0.0014 to 0.0014) | 0.9568 | 9364 |
| FluidIntelligence2 | -0.0007 (-0.0021 to 0.0006) | 0.3065 | 5849 | 0.0022 (0.0005 to 0.0038) | 0.0110 | 5502 |
| FluidIntelligence3 | 0.0035 (-0.0012 to 0.0083) | 0.1473 | 523 | 0.0023 (-0.0027 to 0.0073) | 0.3738 | 478 |
| DurationNumericPtrail1Ins2 | 0.0012 (-0.0722 to 0.0745) | 0.9753 | 4418 | -0.0580 (-0.1520 to 0.0359) | 0.2260 | 4128 |
| **DurationNumericPtrail1Ins3** | -0.0988 (-0.2754 to 0.0778) | 0.2735 | 522 | **0.3879 (0.0064 to 0.7694)** | **0.0469** | **476** |
| **DurationAlphaNumePtrail2Ins2** | **0.2668 (0.0498 to 0.4839)** | **0.0160** | **4304** | **-0.3074 (-0.5947 to -0.0201)** | **0.0360** | **3998** |
| DurationAlphaNumePtrail2Ins3 | 0.2329 (-0.4697 to 0.9355) | 0.5162 | 511 | -0.0710 (-0.8519 to 0.7099) | 0.8586 | 465 |
| NumberWordPairsCorrectAssociated2 | -0.0017 (-0.0037 to 0.0003) | 0.1058 | 4474 | 0.0015 (-0.0011 to 0.0041) | 0.2510 | 4172 |
| NumberWordPairsCorrectAssociated3 | 0.0022 (-0.0039 to 0.0083) | 0.4742 | 534 | 0.0041 (-0.0031 to 0.0113) | 0.2629 | 487 |
| MaxDigitRememCorrectly0 | -0.0009 (-0.0022 to 0.0004) | 0.1698 | 2422 | 0.0006 (-0.0011 to 0.0023) | 0.4895 | 2651 |
| MaxDigitRememCorrectly2 | -0.0009 (-0.0020 to 0.0003) | 0.1356 | 3396 | -0.0002 (-0.0016 to 0.0013) | 0.8079 | 3221 |
| MaxDigitRememCorrectly3 | 0.0001 (-0.0034 to 0.0036) | 0.9412 | 390 | -0.0002 (-0.0041 to 0.0037) | 0.9107 | 361 |
| ProspectiveMemR0 | -0.0007 (-0.0024 to 0.0009) | 0.3903 | 8378 | -0.0008 (-0.0028 to 0.0012) | 0.4197 | 9468 |
| ProspectiveMemR1 | -0.0003 (-0.0034 to 0.0030) | 0.8613 | 2294 | 0.0002 (-0.0044 to 0.0050) | 0.9445 | 2271 |
| ProspectiveMemR3 | 0.0023 (-0.0045 to 0.0093) | 0.5209 | 534 | 0.0036 (-0.0061 to 0.0142) | 0.4805 | 487 |

*Table S. 15 Associations between MVPA and cognitive function using fully model (categorical MVPA) stratified by sex. Bold results show significant (p<0.05) results.*

|  | Males | | | Females | | |
|---|---|---|---|---|---|---|
| Cognitive_Var | Estimate_CI | P_Value | N_Obs | Estimate_CI | P_Value | N_Obs |
| ReactionTime0 | -0.3819 (-3.6682 to 2.9045) | 0.8198 | 21464 | 0.1655 (-2.7547 to 3.0857) | 0.9116 | 24322 |
| ReactionTime2 | 0.0211 (-6.9604 to 7.0025) | 0.9953 | 5930 | -5.3871 (-11.7068 to 0.9327) | 0.0948 | 5590 |
| ReactionTime3 | -7.7586 (-29.8373 to 14.3201) | 0.4913 | 525 | -6.1133 (-29.0245 to 16.7978) | 0.6012 | 477 |
| FluidIntelligence0 | 0.0981 (-0.0002 to 0.1963) | 0.0505 | 8287 | -0.0039 (-0.0825 to 0.0746) | 0.9220 | 9364 |
| FluidIntelligence2 | -0.0753 (-0.1971 to 0.0466) | 0.2260 | 5849 | 0.0516 (-0.0533 to 0.1565) | 0.3348 | 5502 |
| FluidIntelligence3 | 0.1539 (-0.2482 to 0.5559) | 0.4536 | 523 | 0.2113 (-0.1434 to 0.5659) | 0.2435 | 478 |
| **DurationNumericPtrail1Ins2** | -0.9679 (-7.6708 to 5.7350) | 0.7772 | 4418 | **6.3018 (0.2898 to 12.3138)** | **0.0400** | **4128** |
| DurationNumericPtrail1Ins3 | -11.0674 (-25.8182 to 3.6834) | 0.1420 | 522 | 6.9164 (-20.1395 to 33.9722) | 0.6166 | 476 |
| DurationAlphaNumePtrail2Ins2 | 6.5167 (-13.4618 to 26.4953) | 0.5227 | 4304 | -5.4624 (-23.9206 to 12.9959) | 0.5619 | 3998 |
| DurationAlphaNumePtrail2Ins3 | -12.0260 (-70.8636 to 46.8115) | 0.6889 | 511 | 3.6293 (-51.4857 to 58.7442) | 0.8974 | 465 |
| NumberWordPairsCorrectAssociated2 | 0.0963 (-0.0868 to 0.2794) | 0.3025 | 4474 | 0.1031 (-0.0638 to 0.2700) | 0.2260 | 4172 |
| NumberWordPairsCorrectAssociated3 | 0.1564 (-0.3598 to 0.6725) | 0.5529 | 534 | **0.6180 (0.1145 to 1.1215)** | **0.0165** | **487** |
| MaxDigitRememCorrectly0 | 0.0284 (-0.0861 to 0.1429) | 0.6269 | 2422 | 0.0688 (-0.0292 to 0.1668) | 0.1690 | 2651 |
| MaxDigitRememCorrectly2 | -0.0282 (-0.1322 to 0.0757) | 0.5944 | 3396 | -0.0452 (-0.1378 to 0.0473) | 0.3384 | 3221 |

| | | | | | | |
|---|---|---|---|---|---|---|
| MaxDigitRememCorrectly3 | 0.1325 (-0.1677 to 0.4327) | 0.3876 | 390 | 0.0398 (-0.2287 to 0.3084) | 0.7714 | 361 |
| ProspectiveMemR0 | 0.0113 (-0.1261 to 0.1472) | 0.8714 | 8378 | -0.0364 (-0.1505 to 0.0773) | 0.5307 | 9468 |
| ProspectiveMemR1 | 0.0153 (-0.2672 to 0.2914) | 0.9146 | 2294 | -0.0100 (-0.2807 to 0.2587) | 0.9423 | 2271 |
| **ProspectiveMemR3** | **0.5884 (0.0615 to 1.1040)** | **0.0265** | **534** | -0.0098 (-0.6539 to 0.6183) | 0.9758 | 487 |

*Table S. 16 Associations between MVPA and cognitive function using fully model (MVPA in minutes per week) stratified by obesity status. Bold results show significant (p<0.05) results.*

| | Non-obese | | | Obese | | |
|---|---|---|---|---|---|---|
| Cognitive_Var | Estimate_CI | P_Value | N_Obs | Estimate_CI | P_Value | N_Obs |
| ReactionTime0 | 0.0059 (-0.0272 to 0.0390) | 0.7279 | 36743 | 0.0548 (-0.0280 to 0.1375) | 0.1947 | 9041 |
| ReactionTime2 | 0.0483 (-0.0168 to 0.1135) | 0.1460 | 9522 | 0.0592 (-0.1123 to 0.2307) | 0.4988 | 1998 |
| ReactionTime3 | 0.1223 (-0.0993 to 0.3439) | 0.2799 | 830 | 0.5259 (-0.0051 to 1.0568) | 0.0541 | 172 |
| FluidIntelligence0 | 0.0004 (-0.0006 to 0.0013) | 0.4491 | 14006 | -0.0005 (-0.0027 to 0.0017) | 0.6494 | 3645 |
| FluidIntelligence2 | 0.0004 (-0.0007 to 0.0015) | 0.5193 | 9386 | 0.0002 (-0.0028 to 0.0031) | 0.9144 | 1965 |
| FluidIntelligence3 | 0.0026 (-0.0010 to 0.0063) | 0.1523 | 829 | 0.0073 (-0.0033 to 0.0179) | 0.1790 | 172 |
| DurationNumericPtrail1Ins2 | -0.0074 (-0.0682 to 0.0534) | 0.8121 | 7082 | -0.0383 (-0.2061 to 0.1295) | 0.6548 | 1464 |
| DurationNumericPtrail1Ins3 | 0.1549 (-0.0691 to 0.3789) | 0.1756 | 827 | -0.1984 (-0.6856 to 0.2887) | 0.4247 | 1422 |
| DurationAlphaNumePtrail2Ins2 | 0.0812 (-0.1034 to 0.2657) | 0.3886 | 6880 | 0.3697 (-1.1262 to 1.8656) | 0.6288 | 169 |
| DurationAlphaNumePtrail2Ins3 | 0.0683 (-0.4837 to 0.6202) | 0.8085 | 807 | 0.0003 (-0.0041 to 0.0047) | 0.8892 | 1476 |
| NumberWordPairsCorrectAssociated2 | -0.0004 (-0.0021 to 0.0013) | 0.6477 | 7170 | 0.0104 (-0.0023 to 0.0230) | 0.1104 | 174 |
| NumberWordPairsCorrectAssociated3 | 0.0022 (-0.0027 to 0.0072) | 0.3774 | 847 | -0.0016 (-0.0042 to 0.0010) | 0.2389 | 1034 |
| MaxDigitRememCorrectly0 | -0.0000 (-0.0011 to 0.0011) | 0.9655 | 4039 | -0.0002 (-0.0027 to 0.0024) | 0.9058 | 1127 |
| MaxDigitRememCorrectly2 | -0.0007 (-0.0016 to 0.0003) | 0.1631 | 5490 | 0.0062 (-0.0019 to 0.0142) | 0.1372 | 124 |
| MaxDigitRememCorrectly3 | -0.0007 (-0.0034 to 0.0020) | 0.6171 | 627 | -0.0011 (-0.0042 to 0.0021) | 0.4809 | 3688 |
| ProspectiveMemR0 | -0.0006 (-0.0020 to 0.0008) | 0.3894 | 14158 | -0.0058 (-0.0127 to 0.0015) | 0.1084 | 889 |
| ProspectiveMemR1 | 0.0008 (-0.0020 to 0.0037) | 0.5876 | 3676 | 0.0007 (-0.0032 to 0.0047) | 0.7296 | 2010 |
| ProspectiveMemR3 | 0.0016 (-0.0040 to 0.0075) | 0.5872 | 847 | 0.0160 (-0.0035 to 0.0395) | 0.1405 | 174 |

*Table S. 17 Associations between MVPA and cognitive function using fully model (categorical MVPA in minutes per week) stratified by obesity status. Bold results show significant (p<0.05) results.*

| | Non-obese | | | Obese | | |
|---|---|---|---|---|---|---|
| Cognitive_Var | Estimate_CI | P_Value | N_Obs | Estimate_CI | P_Value | N_Obs |
| ReactionTime0 | -0.3869 (-2.8286 to 2.0548) | 0.7561 | 36744 | 2.8704 (-1.8562 to 7.5970) | 0.2340 | 9042 |
| ReactionTime2 | -3.9583 (-9.2014 to 1.2847) | 0.1390 | 9522 | 3.3526 (-6.8383 to 13.5435) | 0.5191 | 1998 |
| ReactionTime3 | -12.4982 (-30.6029 to 5.6065) | 0.1764 | 830 | 15.1488 (-16.4972 to 46.7948) | 0.3496 | 172 |
| FluidIntelligence0 | 0.0052 (-0.0647 to 0.0751) | 0.8839 | 14006 | 0.1163 (-0.0118 to 0.2444) | 0.0752 | 3645 |
| FluidIntelligence2 | 0.0034 (-0.0860 to 0.0928) | 0.9413 | 9386 | -0.0602 (-0.2350 to 0.1146) | 0.4998 | 1965 |
| FluidIntelligence3 | 0.1314 (-0.1646 to 0.4274) | 0.3845 | 829 | 0.3474 (-0.2798 to 0.9745) | 0.2794 | 172 |
| **DurationNumericPtrail1Ins2** | **5.9076 (0.9411 to 10.8740)** | **0.0198** | **7082** | -3.3377 (-13.4489 to 6.7734) | 0.5177 | 1464 |

| Cognitive_Var | Estimate_CI | P_Value | N_Obs | Estimate_CI | P_Value | N_Obs |
|---|---|---|---|---|---|---|
| DurationNumericPtrail1Ins3 | -1.7814 (-20.0680 to 16.5052) | 0.8486 | 827 | -1.5602 (-31.1232 to 28.0028) | 0.9176 | 1422 |
| DurationAlphaNumePtrail2Ins2 | 0.1850 (-14.9746 to 15.3446) | 0.9809 | 6880 | 26.4079 (-61.4952 to 114.3109) | 0.5569 | 169 |
| DurationAlphaNumePtrail2Ins3 | -6.3943 (-51.4993 to 38.7106) | 0.7812 | 807 | 0.0411 (-0.2247 to 0.3070) | 0.7617 | 1476 |
| NumberWordPairsCorrectAssociated2 | 0.1211 (-0.0172 to 0.2594) | 0.0862 | 7170 | 0.2210 (-0.5318 to 0.9738) | 0.5658 | 174 |
| NumberWordPairsCorrectAssociated3 | 0.4577 (0.0519 to 0.8634) | 0.0273 | 847 | -0.1333 (-0.2879 to 0.0213) | 0.0914 | 1034 |
| MaxDigitRememCorrectly0 | 0.1141 (0.0300 to 0.1982) | 0.0079 | 4039 | -0.0280 (-0.1835 to 0.1275) | 0.7239 | 1127 |
| MaxDigitRememCorrectly2 | -0.0428 (-0.1198 to 0.0342) | 0.2757 | 5490 | 0.2144 (-0.2819 to 0.7107) | 0.3991 | 124 |
| MaxDigitRememCorrectly3 | 0.0778 (-0.1422 to 0.2978) | 0.4883 | 627 | -0.0313 (-0.2141 to 0.1517) | 0.7373 | 3688 |
| ProspectiveMemR0 | -0.0129 (-0.1120 to 0.0856) | 0.7978 | 14158 | -0.3695 (-0.7978 to 0.0547) | 0.0887 | 889 |
| ProspectiveMemR1 | 0.0991 (-0.1175 to 0.3127) | 0.3660 | 3676 | -0.0944 (-0.3237 to 0.1339) | 0.4185 | 2010 |
| ProspectiveMemR3 | 0.2289 (-0.2190 to 0.6641) | 0.3084 | 847 | 0.9884 (-0.0239 to 2.0566) | 0.0602 | 174 |

*Table S. 18 Associations between MVPA and subcortical brain region using fully model (MVPA in minutes per week and categorical MVPA) stratified by age for participants under 65 years. Bold results show significant (p<0.05) results.*

| Fully adjusted | MVPA per day in Mins | | MVPA per WHO guidelines | | |
|---|---|---|---|---|---|
| Cognitive_Var | Estimate_CI | P_Value | Estimate_CI | P_Value | N_Obs |
| **VolumeHippocampusInst2** | **0.4688 (0.2077 to 0.7299)** | **0.00043** | **32.6001 (12.7303 to 52.4699)** | **0.0013** | 8524 |
| VolumeHippocampusInst3 | 0.0938 (-0.6209 to 0.8084) | 0.7971 | 9.4478 (-45.6931 to 64.5887) | 0.7370 | 1117 |
| VolumeAccumbensInst2 | 0.0336 (-0.0280 to 0.0951) | 0.2851 | 2.8340 (-1.8475 to 7.5154) | 0.2354 | 8524 |
| VolumeAccumbensInst3 | -0.0067 (-0.1711 to 0.1578) | 0.9365 | 3.4395 (-9.2493 to 16.1282) | 0.5953 | 1117 |
| **VolumeAmygdalaInst2** | **0.2204 (0.0878 to 0.3529)** | **0.0011** | **14.1776 (4.0897 to 24.2654)** | **0.0058** | 8524 |
| VolumeAmygdalaInst3 | 0.1682 (-0.1837 to 0.5201) | 0.3490 | 20.1270 (-7.0075 to 47.2615) | 0.1462 | 1117 |
| **VolumeCaudateInst2** | **0.3839 (0.1286 to 0.6392)** | **0.0032** | **23.5103 (4.0822 to 42.9384)** | **0.0177** | 8524 |
| VolumeCaudateInst3 | 0.2543 (-0.4074 to 0.9159) | 0.4514 | 5.1623 (-45.9023 to 56.2269) | 0.8429 | 1117 |
| **VolumePallidumInst2** | **0.3284 (0.1961 to 0.4607)** | **1.16E-06** | **20.5206 (10.4502 to 30.5910)** | **6.56E-05** | 8524 |
| VolumePallidumInst3 | 0.1477 (-0.2238 to 0.5192) | 0.4359 | 23.4999 (-5.1412 to 52.1411) | 0.1080 | 1117 |
| **VolumePutamenInst2** | **0.4278 (0.1133 to 0.7423)** | **0.0076** | **37.1765 (13.2472 to 61.1059)** | **0.0023** | 8524 |
| **VolumePutamenInst3** | 0.5620 (-0.2732 to 1.3971) | 0.1874 | **65.4109 (1.0345 to 129.7873)** | **0.0466** | **1117** |
| VolumeThalamusInst2 | 0.3578 (-0.0568 to 0.7725) | 0.0908 | 42.5474 (11.0020 to 74.0928) | 0.0082 | 8524 |
| VolumeThalamusInst3 | 0.3150 (-0.7676 to 1.3976) | 0.5686 | 34.4870 (-49.0338 to 118.0077) | 0.4185 | 1117 |

*Table S. 19 Associations between MVPA and subcortical brain region using fully model (MVPA in minutes per week and categorical MVPA) stratified by age for participants over 65 years. Bold results show significant (p<0.05) results.*

| | MVPA per day in Mins | | MVPA per WHO guidelines | | |
|---|---|---|---|---|---|
| Cognitive_Var | Estimate_CI | P_Value | Variable | Estimate_CI | N_Obs |
| VolumeHippocampusInst2 | 0.2408 (-0.1132 to 0.5949) | 0.18253 | 11.1062 (-14.7395 to 36.9519) | 0.3997 | 5412 |
| VolumeHippocampusInst3 | -0.3181 (-1.5509 to 0.9148) | 0.6133 | -7.6220 (-98.6758 to 83.4318) | 0.8697 | 427 |
| **VolumeAccumbensInst2** | **0.1574 (0.0779 to 0.2369)** | **0.0001** | **9.2425 (3.4356 to 15.0494)** | **0.0018** | 5412 |
| VolumeAccumbensInst3 | 0.1387 (-0.1258 to 0.4032) | 0.3045 | 3.6483 (-15.8998 to 23.1964) | 0.7147 | 427 |
| VolumeAmygdalaInst2 | 0.0719 (-0.1013 to 0.2451) | 0.4159 | 9.8023 (-2.8385 to 22.4431) | 0.1286 | 5412 |
| VolumeAmygdalaInst3 | 0.4093 (-0.1688 to 0.9874) | 0.1660 | 18.3323 (-24.4174 to 61.0819) | 0.4011 | 427 |
| VolumeCaudateInst2 | 0.1255 (-0.2059 to 0.4568) | 0.4580 | 18.5700 (-5.6129 to 42.7528) | 0.1323 | 5412 |

| | | | |
|---|---|---|---|---|---|
| VolumeCaudateInst3 | 0.0509 (-1.0657 to 1.1675) | 0.9288 | -53.1010 (-135.3842 to 29.1822) | 0.2066 | 427 |
| **VolumePallidumInst2** | **0.2888 (0.0932 to 0.4844)** | **0.0038** | **21.3282 (7.0520 to 35.6044)** | **0.0034** | **5412** |
| VolumePallidumInst3 | 0.2829 (-0.3989 to 0.9648) | 0.4164 | -1.6650 (-52.0471 to 48.7170) | 0.9483 | 427 |
| VolumePutamenInst2 | 0.4087 (-0.0082 to 0.8255) | 0.0547 | 20.1777 (-10.2544 to 50.6098) | 0.1938 | 5412 |
| VolumePutamenInst3 | 0.7354 (-0.7267 to 2.1976) | 0.3248 | -33.8732 (-141.9103 to 74.1640) | 0.5392 | 427 |
| **VolumeThalamusInst2** | **0.6784 (0.1663 to 1.1905)** | **0.0094** | 29.4645 (-7.9298 to 66.8588) | 0.1225 | 5412 |
| VolumeThalamusInst3 | 0.9416 (-0.7905 to 2.6737) | 0.2872 | -55.9002 (-183.8491 to 72.0487) | 0.3923 | 427 |

*Table S. 20 Associations between MVPA and subcortical brain region using fully model (MVPA in minutes per week) stratified by sex. Bold results show significant (p<0.05) results.*

| | Males – fully adjusted | | | Females – fully adjusted | | |
|---|---|---|---|---|---|---|
| Cognitive_Var | Estimate_CI | P_Value | N_Obs | Estimate_CI | P_value | N_Obs |
| VolumeHippocampusInst2 | -0.0218 (-0.3936 to 0.3500) | 0.9086 | 4170 | 0.1983 (-0.1905 to 0.5871) | 0.31747 | 4217 |
| VolumeHippocampusInst3 | -0.4246 (-1.7370 to 0.8878) | 0.5264 | 370 | -0.0658 (-1.2707 to 1.1391) | 0.91481 | 375 |
| VolumeAccumbensInst2 | 0.1132 (0.0296 to 0.1969) | 0.0080 | 4170 | 0.0345 (-0.0559 to 0.1249) | 0.45466 | 4217 |
| VolumeAccumbensInst3 | 0.0095 (-0.2812 to 0.3003) | 0.9487 | 370 | 0.1624 (-0.1017 to 0.4266) | 0.22898 | 375 |
| VolumeAmygdalaInst2 | 0.0066 (-0.1815 to 0.1948) | 0.9448 | 4170 | **0.2096 (0.0124 to 0.4069)** | **0.03733** | **4217** |
| VolumeAmygdalaInst3 | 0.4110 (-0.2581 to 1.0801) | 0.2293 | 370 | 0.2099 (-0.3523 to 0.7721) | 0.46482 | 375 |
| VolumeCaudateInst2 | 0.3096 (-0.0427 to 0.6618) | 0.0850 | 4170 | 0.2402 (-0.1478 to 0.6283) | 0.2250 | 4217 |
| VolumeCaudateInst3 | 0.6293 (-0.5101 to 1.7686) | 0.2797 | 370 | 0.1457 (-1.0327 to 1.3240) | 0.80871 | 375 |
| **VolumePallidumInst2** | **0.2631 (0.0613 to 0.4649)** | **0.0106** | **4170** | **0.3485 (0.1229 to 0.5740)** | **0.0024** | **4217** |
| VolumePallidumInst3 | 0.2859 (-0.4526 to 1.0243) | 0.4484 | 370 | 0.0882 (-0.5608 to 0.7371) | 0.7901 | 375 |
| VolumePutamenInst2 | 0.4035 (-0.0318 to 0.8388) | 0.0693 | 4170 | 0.2938 (-0.1772 to 0.7648) | 0.2215 | 4217 |
| **VolumePutamenInst3** | **1.6000 (0.1475 to 3.0526)** | **0.0315** | **370** | 0.2484 (-1.2166 to 1.7135) | 0.7398 | 375 |
| **VolumeThalamusInst2** | 0.4457 (-0.0889 to 0.9803) | 0.1023 | 4170 | **0.6379 (0.0409 to 1.2349)** | **0.0362** | **4217** |
| VolumeThalamusInst3 | 0.4366 (-1.3672 to 2.2404) | 0.6355 | 370 | 0.1147 (-1.5893 to 1.8186) | 0.89513 | 375 |

*Table S. 21 Associations between MVPA and subcortical brain region using fully model (categorical MVPA) stratified by sex. Bold results show significant (p<0.05) results.*

| Cognitive_Var | Males – fully adjusted | | | Females – fully adjusted | | |
|---|---|---|---|---|---|---|
| | Estimate_CI | P_Value | N_Obs | Estimate_CI | P_Value | N_Obs |
| VolumeHippocampusInst2 | 15.7103 (-17.1806 to 48.6012) | 0.3492 | 4170 | 0.7677 (-23.7863 to 25.3216) | 0.9511 | 4217 |
| VolumeHippocampusInst3 | -73.4399 (-182.4099 to 35.5301) | 0.1874 | 370 | 26.8310 (-59.1851 to 112.8471) | 0.5413 | 375 |
| VolumeAccumbensInst2 | 7.1030 (-0.3018 to 14.5078) | 0.0602 | 4170 | 2.5029 (-3.2044 to 8.2103) | 0.3901 | 4217 |
| VolumeAccumbensInst3 | -6.1359 (-30.3188 to 18.0469) | 0.6193 | 370 | 12.5125 (-6.3491 to 31.3742) | 0.1944 | 375 |
| **VolumeAmygdalaInst2** | 5.1553 (-11.4877 to 21.7983) | 0.5438 | 4170 | **13.9965 (1.5398 to 26.4531)** | **0.0277** | **4217** |
| VolumeAmygdalaInst3 | 46.9542 (-8.6049 to 102.5133) | 0.0985 | 370 | 33.7514 (-6.2818 to 73.7846) | 0.0993 | 375 |
| VolumeCaudateInst2 | 33.1998 (2.0401 to 64.3596) | 0.0368 | 4170 | 10.1712 (-14.3365 to 34.6789) | 0.4160 | 4217 |

| | | | | | | |
|---|---|---|---|---|---|---|
| VolumeCaudateInst3 | -55.1224 (-149.8881 to 39.6432) | 0.2550 | 370 | -27.8775 (-111.9967 to 56.2416) | 0.5164 | 375 |
| **VolumePallidumInst2** | **21.3581 (3.5031 to 39.2131)** | **0.0191** | **4170** | **20.6927 (6.4456 to 34.9399)** | **0.0044** | **4217** |
| VolumePallidumInst3 | 20.0615 (-41.3854 to 81.5084) | 0.5227 | 370 | -8.7997 (-55.1423 to 37.5429) | 0.7100 | 375 |
| VolumePutamenInst2 | 31.4506 (-7.0615 to 69.9628) | 0.1095 | 4170 | 17.3925 (-12.3531 to 47.1381) | 0.2519 | 4217 |
| VolumePutamenInst3 | 44.3260 (-77.2278 to 165.8799) | 0.4753 | 370 | -29.0918 (-133.7023 to 75.5188) | 0.5861 | 375 |
| VolumeThalamusInst2 | 38.0844 (-9.2128 to 85.3815) | 0.1146 | 4170 | 22.3542 (-15.3589 to 60.0672) | 0.2454 | 4217 |
| VolumeThalamusInst3 | -48.0193 (-198.0445 to 102.0059) | 0.5308 | 370 | -52.6980 (-174.2801 to 68.8841) | 0.3962 | 375 |

*Table S. 22 Associations between MVPA and subcortical brain region using fully model (MVPA in minutes per week) stratified by obesity status. Bold results show significant (p<0.05) results.*

| | Non _obese– fully adjusted | | | Obese Population | | |
|---|---|---|---|---|---|---|
| Cognitive_Var | Estimate_CI | P_Value | N_Obs | Estimate_CI | P_Value | N_Obs |
| VolumeHippocampusInst2 | 0.0941 (-0.1908 to 0.3790) | 0.5174 | 6953 | -0.1882 (-0.9311 to 0.5546) | 0.6195 | 1434 |
| VolumeHippocampusInst3 | -0.3256 (-1.2897 to 0.6385) | 0.5083 | 629 | 1.0938 (-1.3751 to 3.5627) | 0.3874 | 116 |
| **VolumeAccumbensInst2** | **0.0892 (0.0242 to 0.1542)** | **0.0071** | **6953** | 0.0707 (-0.0986 to 0.2400) | 0.4130 | 1434 |
| VolumeAccumbensInst3 | 0.0364 (-0.1673 to 0.2402) | 0.7262 | 629 | 0.4448 (-0.1863 to 1.0759) | 0.1704 | 116 |
| VolumeAmygdalaInst2 | 0.0617 (-0.0820 to 0.2053) | 0.4002 | 6953 | 0.1624 (-0.2219 to 0.5466) | 0.4077 | 1434 |
| VolumeAmygdalaInst3 | 0.2867 (-0.1646 to 0.7379) | 0.2135 | 629 | 0.5232 (-0.9545 to 2.0008) | 0.4894 | 116 |
| **VolumeCaudateInst2** | 0.1645 (-0.1095 to 0.4384) | 0.2393 | 6953 | **0.9926 (0.2456 to 1.7397)** | **0.0093** | **1434** |
| VolumeCaudateInst3 | 0.3589 (-0.5033 to 1.2210) | 0.4149 | 629 | 0.5156 (-2.1991 to 3.2303) | 0.7105 | 116 |
| **VolumePallidumInst2** | **0.3021 (0.1417 to 0.4625)** | **0.0002** | **6953** | 0.3104 (-0.0863 to 0.7071) | 0.1254 | 1434 |
| VolumePallidumInst3 | 0.1387 (-0.3657 to 0.6431) | 0.5902 | 629 | 0.7800 (-0.9562 to 2.5162) | 0.3808 | 116 |
| **VolumePutamenInst2** | **0.3562 (0.0205 to 0.6919)** | **0.0376** | **6953** | 0.4960 (-0.4185 to 1.4105) | 0.2880 | 1434 |
| VolumePutamenInst3 | 0.4809 (-0.5863 to 1.5480) | 0.3775 | 629 | 3.2533 (-0.1207 to 6.6274) | 0.0618 | 116 |
| **VolumeThalamusInst2** | **0.5264 (0.1061 to 0.9467)** | **0.0141** | **6953** | 0.4274 (-0.6966 to 1.5513) | 0.4562 | 1434 |
| VolumeThalamusInst3 | 0.2350 (-1.0640 to 1.5339) | 0.7231 | 629 | 0.8313 (-3.2631 to 4.9258) | 0.6916 | 116 |

*Table S. 23 Associations between MVPA and subcortical brain region using fully model (categorical MVPA) stratified by obesity status. Bold results show significant (p<0.05) results.*

| Cognitive_Var | Non _obese– fully adjusted | | | Obese Population | | |
|---|---|---|---|---|---|---|
| | Estimate_CI | P_Value | N_Obs | Estimate_CI | P_Value | N_Obs |
| VolumeHippocampusInst2 | 6.8791 (-15.6729 to 29.4311) | 0.5500 | 6953 | -0.1882 (-0.9311 to 0.5546) | 0.6195 | 1434 |
| VolumeHippocampusInst3 | -43.0271 (-121.0008 to 34.9465) | 0.2799 | 629 | 1.0938 (-1.3751 to 3.5627) | 0.3874 | 116 |
| VolumeAccumbensInst2 | 3.7970 (-1.3479 to 8.9420) | 0.1481 | 6953 | 0.0707 (-0.0986 to 0.2400) | 0.4130 | 1434 |
| VolumeAccumbensInst3 | -3.5244 (-20.0115 to 12.9626) | 0.6754 | 629 | 0.4448 (-0.1863 to 1.0759) | 0.1704 | 116 |
| VolumeAmygdalaInst2 | 9.7617 (-1.6073 to 21.1307) | 0.0924 | 6953 | 0.1624 (-0.2219 to 0.5466) | 0.4077 | 1434 |
| VolumeAmygdalaInst3 | 24.2834 (-12.2291 to 60.7958) | 0.1929 | 629 | 0.5232 (-0.9545 to 2.0008) | 0.4894 | 116 |
| VolumeCaudateInst2 | 12.8521 (-8.8326 to 34.5368) | 0.2454 | 6953 | 0.9926 (0.2456 to 1.7397) | 0.0093 | 1434 |
| VolumeCaudateInst3 | -51.3884 (-121.0763 to 18.2996) | 0.1489 | 629 | 0.5156 (-2.1991 to 3.2303) | 0.7105 | 116 |
| **VolumePallidumInst2** | **24.3013 (11.6048 to 36.9978)** | **0.0002** | **6953** | 0.3104 (-0.0863 to 0.7071) | 0.1254 | 1434 |

| | | | | | | |
|---|---|---|---|---|---|---|
| VolumePallidumInst3 | -18.3619 (-59.1647 to 22.4408) | 0.3781 | 629 | 0.7800 (-0.9562 to 2.5162) | 0.3808 | 116 |
| VolumePutamenInst2 | 19.8981 (-6.6808 to 46.4771) | 0.1423 | 6953 | 0.4960 (-0.4185 to 1.4105) | 0.2880 | 1434 |
| VolumePutamenInst3 | -49.4163 (-135.7394 to 36.9067) | 0.2623 | 629 | 3.2533 (-0.1207 to 6.6274) | 0.0618 | 116 |
| VolumeThalamusInst2 | 25.7427 (-7.5369 to 59.0222) | 0.1295 | 6953 | 0.4274 (-0.6966 to 1.5513) | 0.4562 | 1434 |
| VolumeThalamusInst3 | -83.3537 (-188.2716 to 21.5643) | 0.1200 | 629 | 0.8313 (-3.2631 to 4.9258) | 0.6916 | 116 |



\end{document}